\documentclass{jfm}
\usepackage{graphicx}
\usepackage{epstopdf, epsfig}
\usepackage{amsmath}
\usepackage{subfig}
\usepackage{xcolor}
\usepackage{soul}
\usepackage{enumitem}

\shorttitle{Modified Snaking}
\shortauthor{S. Azimi and T. M. Schneider}

\title{Modified snaking in plane Couette flow with wall-normal suction}

\author{Sajjad Azimi\aff{1}
 \and Tobias M.  Schneider\aff{1}
  \corresp{\email{tobias.schneider@epfl.ch}}}

\affiliation{\aff{1}Emergent Complexity in Physical Systems Laboratory (ECPS), \'Ecole Polythechnique F\'ed\'erale de Lausanne, CH-1015 Lausanne, Switzerland}

\begin{document}

\maketitle

\begin{abstract}
A specific family of spanwise-localised invariant solutions of plane Couette flow exhibits homoclinic snaking, a process by which spatially localised invariant solutions of a nonlinear partial differential equation smoothly grow additional structure at their fronts while undergoing a sequence of saddle-node bifurcations. 
Homoclinic snaking is well understood in the context of simpler pattern forming systems such as the one-dimensional Swift-Hohenberg equation with  cubic-quintic nonlinearity, whose solutions remarkably well resemble the snaking solutions of plane Couette flow. We study the structural stability of the characteristic snakes-and-ladders structure associated with homoclinic snaking for flow modifications that break symmetries of plane Couette flow. We demonstrate that wall-normal suction modifies the bifurcation structure of three-dimensional plane Couette solutions in the same way, a symmetry-breaking quadratic term modifies solutions of the one-dimensional Swift-Hohenberg equation. These modifications are related to the breaking of the discrete rotational symmetry. At large amplitudes of the symmetry-breaking wall-normal suction the connected snakes-and-ladders structure is destroyed. Previously unknown solution branches are created and can be parametrically continued to vanishing suction. This yields new localised solutions of plane Couette flow that exist in a wide range of Reynolds number. 
\end{abstract}

\begin{keywords}

\end{keywords}

\section{Introduction}\label{intro}
Invariant solutions of the Navier-Stokes equations play a key role for the dynamics of transitional shear flows \citep{Kawahara2012}. These solutions, in the form of equilibria, travelling waves and periodic orbits, have been computed for many canonical shear flows including pipe flow \citep{Faisst2003}, plane Couette flow \citep{Gibson2009}, plane Poisuille flow \citep{Waleffe2003}, and asymptotic suction boundary layer flow \citep{Kreilos2013a}. Invariant solutions are mostly found in small periodic domains, or `minimal flow units' \citep{jimenez1991}. Later investigations have considered extended domains and identified localised invariant solutions that capture large scale flow patterns like turbulent spots, stripes and puffs \citep{Avila2013, Brand2014, Reetz2019a, Reetz2019g}.

The first family of spatially localised invariant solutions in shear flows were calculated by \cite{Schneider2010a} in plane Couette flow. This includes equilibria and travelling waves, which are equilibria in a frame of reference moving relative to the lab frame. These localised invariant solutions in plane Couette flow are specifically noteworthy because they exhibit the characteristic behaviour of homoclinic snaking \citep[see review by][]{Knobloch2015} under parametric continuation \citep{Schneider2010}. Homoclinic snaking is a process previously observed in many dissipative pattern-forming systems such as binary-fluid convection systems \citep{Batiste2006} and optical systems \citep{Firth2007a}, by which localised solutions grow in one direction while undergoing a sequence of successive saddle-node bifurcations \citep{Woods1999}. Homoclinic snaking manifests itself by a characteristic snakes-and-ladders structure in the bifurcation diagram \citep{Burke2007}.

A well-studied one-dimensional model system which supports localized solutions that exhibit homoclinic snaking is the Swift-Hohenberg equation $\partial_t u=r u + (1+\partial^2_x)^2 u+\mathcal{N}(u)$, for a real-valued function $u(x)$ on the real axis with $r$ the bifurcation parameter and a nonlinearity $\mathcal{N}(u)$ \citep{Burke2006, Burke2007, Beck2009, Knobloch2019}. Several variants of the Swift-Hohenberg equation with differing forms of the nonlinear term have been considered. Most studied are both a quadratic-cubic $\mathcal{N}=b_2 u^2 - u^3$ and a cubic-quintic $\mathcal{N} = b_3 u^3 - u^5$ nonlinearity, where $b_2$ and $b_3$ are adjustable parameters. For both nonlinear terms, the Swift-Hohenberg equation supports localised solutions arranged in a snaking bifurcation structure. The localised invariant solutions of the Navier-Stokes equations in plane Couette geometry share remarkably similar properties with solutions of the Swift-Hohenberg equation with the cubic-quintic nonlinearity \citep{Schneider2010}:\\
(i) First, the bifurcation diagram with its characteristic snakes-and-ladders structure is almost indistinguishable from that of the Swift-Hohenberg equation.\\ 
(ii) Second, the three-dimensional snaking solutions of the Navier-Stokes equations very closely resemble one-dimensional snaking solutions of Swift-Hohenberg, when three-dimensional velocity fields are averaged in the downstream direction and zero-sets of downstream velocity are visualised as a function of the spanwise coordinate.\\
The almost perfect resemblance of three-dimensional  Navier-Stokes and one-dimensional Swift-Hohenberg solutions is observed for the downstream wavelength of $4 \pi$ studied in \cite{Schneider2010}. Subtle modifications related to internal deformations of the three-dimensional flow field appear when the downstream wavelength changes, but the characteristic snakes-and-ladders structure remains intact \citep{Gibson2016}. 
Why localised solutions of the three-dimensional Navier-Stokes equations closely resemble solutions of the one-dimensional Swift-Hohenberg equation and are organized in a snakes-and-ladders bifurcation structure is not fully understood. The detailed analysis of model systems including Swift-Hohenberg using concepts such as spatial dynamics indicates the importance of discrete symmetries for an equation to support homoclinic snaking \citep{Champneys1998, Burke2009, Knobloch2015}. In both the Navier-Stokes solutions and in solutions of the Swift-Hohenberg equation with cubic-quintic nonlinearity, the snakes-and-ladders bifurcation structure is composed of two pairs of intertwined snaking branches along which the solutions are invariant under discrete symmetries that are part of the equivariance group of the respective system. The snaking branches of symmetric solutions are connected by so-called rungs which emerge in symmetry-breaking pitchfork bifurcations. The rungs do not possess discrete symmetries and together with the snaking branches form the snakes-and-ladders structure.

To investigate the importance of discrete symmetries for the snaking structure of the Swift-Hohenberg equation with qubic-quintic nonlinearity, \cite{Houghton2011} introduced an additional quadratic term in the equation to break the odd symmetry of the system $R_2 : x\rightarrow -x,\ u\rightarrow -u$. When the amplitude $\epsilon$ of this symmetry-breaking term $\epsilon u^2$ is increased from zero, the bifurcation structure changes.
One pair of snaking branches breaks into disconnected pieces while the other pair splits into two distinct curves in an amplitude versus $r$ bifurcation diagram. Those curves are connected by z- and s-shaped non-symmetric solution branches that are composed of solutions that for $\epsilon=0$ form rungs and the snaking branches that disappear when the symmetry is broken.

In the Swift-Hohenberg equation with cubic-quintic nonlinearity, breaking a specific discrete symmetry destroys the snakes-and-ladders bifurcation structure that solutions of Navier-Stokes in plane Couette geometry remarkably well resemble. To investigate the significance of discrete symmetries of the three-dimensional Navier-Stokes equations, we break a symmetry of plane Couette and study the structural stability of the snakes-and-ladders structure under controlled symmetry breaking. We specifically apply wall-normal suction to break the rotational symmetry of plane Couette flow. At small amplitude, suction modifies the snakes-and-ladders structure of plane Couette flow in the same way, a quadratic term modifies the snakes-and-ladders structure of the Swift-Hohenberg equation with the cubic-quintic nonlinearity. 
At higher values of suction velocity, the snaking branches that remain intact at small suction velocity also break down. The solutions form separated branches which involve solutions that can be followed to zero suction velocity but are not part of the snakes-and-ladders structure of plane Couette flow without suction. We thereby identify previously unknown localised solutions of plane Couette flow that exist in a much wider range of Reynolds numbers than the snaking solutions studied previously. The structural modifications of the bifurcation structure at large amplitude of the symmetry breaking terms is not observed in the Swift-Hohenberg system where increasing the amplitude of the quadratic term transforms the snakes-and-ladders structure into a modified snakes-and-ladders structure similar to the bifurcation structure of the localised solutions of the Swift-Hohenberg equation with quadratic-cubic nonlinearity. Thus, at small suction amplitudes, suction has a similar effect as the symmetry-breaking term within the cubic-quintic Swift-Hohenberg case while at larger suction amplitudes, the bifurcation structure of 3D Navier-Stokes solutions significantly differs from that of the analogous Swift-Hohenberg problem. Both in the one-dimensional model system and in the full three-dimensional Navier-Stokes problem, the initial breakdown of the snakes-and-ladders structure can be explained in terms of symmetry breaking. 

This manuscript is organised as follows: In section \ref{sys_method}, we introduce the plane Couette system with wall-normal suction and discuss its symmetry properties. Section \ref{no_suction} reviews the snaking solutions and the snakes-and-ladders structure of plane Couette flow at zero suction. Section \ref{modified_snaking} presents the key observations on how wall-normal suction modifies the snaking structure. In section \ref{discussion}, features of the modified snaking structure are discussed and related to symmetry properties including symmetry subspaces of the flow. In the final section \ref{conclusion}, results are summarised and an outlook for future work is provided.

\section{System and methodology}\label{sys_method}

Plane Couette flow (PCF) is the flow of a Newtonian fluid between two parallel plates at a distance of $2H$ which move in opposite directions with a constant relative velocity of $2U_w$. The laminar solution of the Navier-Stokes equations in plane Couette flow is a linear profile. We investigate the effect of wall-normal suction on the snaking solutions of plane Couette flow. The wall-normal suction modifies the laminar flow. After nondimensionalisation of the velocities with respect to half of the relative velocity of the plates, $U_w$, and the lengths with respect to half of the gap width between the plates, $H$, the laminar solution takes the form
\begin{align*}
\mathbf{U}(y)=U_x(y)\hat{e}_x-V_s\hat{e}_y,
\end{align*}
where the coordinate system $(x, y, z)$ is aligned with the streamwise $(\hat{e}_x)$, the wall-normal $(\hat{e}_y)$, and the spanwise directions $(\hat{e}_z)$; $\mathbf{U}(y)$ is the laminar solution; $V_s$ is the nondimensionalised wall-normal suction velocity; and
\begin{align*}
U_x(y)=1-\frac{1}{\sinh(-Re V_s)}\left(\exp(-Re V_s)-\exp(-y Re V_s)\right).
\end{align*}
Decomposition of the total velocity into the laminar solution, $\mathbf{U}(y)$ as a base flow, and the deviation from the laminar solution, $\mathbf{u}(x,y,z,t)$, yields the Navier-Stokes equations in perturbative form
\begin{align*}
\frac{\partial \mathbf{u}}{\partial t} + U_x(y)\frac{\partial \mathbf{u}}{\partial x}-V_s\frac{\partial \mathbf{u}}{\partial y}+v\frac{d U_x(y)}{dy}\hat{e}_x+\mathbf{u}\cdot\nabla\mathbf{u}=-\nabla p + \frac{1}{Re}\nabla^2\mathbf{u}
\end{align*}
where the Reynolds number is defined as $Re=UH/\nu$, with $\nu$ the kinematic viscosity of the fluid. The boundary conditions for $\mathbf{u}$ are periodic in streamwise and spanwise directions,
\begin{align*}
\mathbf{u}(-L_x/2,y,z,t) &= \mathbf{u}(L_x/2,y,z,t),\\
\mathbf{u}(x,y,-L_z/2,t) &= \mathbf{u}(x,y,L_z/2,t),
\end{align*}
where $L_x$ and $L_z$ are the length and width of the computational domain in streamwise and spanwise directions, respectively. At the top and bottom plate Dirichlet conditions 
\begin{align*}
\mathbf{u}(x,1,z,t) &= (1,-V_s,0),\\
\mathbf{u}(x,-1,z,t) &= (-1,-V_s,0)
\end{align*}
are satisfied. Zero pressure gradient in both streamwise and spanwise directions is imposed.

Following \cite{Gibson2016}, we present bifurcation diagrams in terms of energy dissipation of the deviation from the laminar solution, normalised by the length of the channel $L_x$. The considered solutions are localised and thus independent of the width $L_z$ of the domain. Consequently, we do not normalise by the width $L_z$. The streamwise averaged energy dissipation for an equilibrium or travelling wave solution equals the energy input rate $I$ and can be expressed as 
\begin{align*}
    D=I=\frac{1}{2L_x}\int_{-L_z/2}^{L_z/2}\int_{-L_x/2}^{L_x/2}\left(\frac{\partial \mathbf{u}}{\partial y}\bigg|_{y=-1}+\frac{\partial \mathbf{u}}{\partial y}\bigg|_{y=1}\right) dx dz
\end{align*}
in units of $U_w^2$. $D$ serves as a measure for the width of a localised solution that approaches laminar flow ($\mathbf{u}=0) $ and only generates nonzero dissipation in the non-laminar part of the solution. 

Plane Couette flow ($V_s=0$) is invariant under continuous translations in $x$ and $z$ directions, discrete reflection in $z$ direction and a discrete rotation of $180^\circ$ around the $z$ axis. The rotation is equivalent to two successive discrete reflections in $x$ and $y$ directions. Following the conventions of \cite{Gibson2008}, symmetries of the governing equations and the boundary conditions of PCF are expressed as
\begin{align*}
\sigma_z : [u,v,w](x,y,z) &\rightarrow[u,v,-w](x,y,-z),\\
\sigma_{xy} : [u,v,w](x,y,z) &\rightarrow[-u,-v,w](-x,-y,z),\\
\tau : [u,v,w](x,y,z) &\rightarrow[u,v,w](x+\delta x,y,z+\delta z),
\end{align*}
where $\delta x$ and $\delta z$ are arbitrary displacements in the streamwise and spanwise directions, respectively. All products of $\sigma_z$, $\sigma_{xy}$ and $\tau$ compose the symmetry, or equivariance, group of plane Couette flow ($V_s=0$).
The equivariance group contains the discrete symmetries the snaking solutions are invariant under. These are the inversion symmetry $\sigma_{xyz}$ and the shift-reflect symmetry $\tau_x\sigma_z$:
\begin{align*}
\sigma_{xyz} : [u,v,w](x,y,z) &\rightarrow[-u,-v,-w](-x,-y,-z),\\
\tau_x\sigma_{z} : [u,v,w](x,y,z) &\rightarrow[u,v,-w](x+L_x/2,y,-z).
\end{align*}
A nonzero wall normal suction velocity breaks the rotational symmetry of the system, $\sigma_{xy}$. Therefore, for plane Couette flow with a nonzero wall-normal suction, the equivariance group consists of all products of the translational symmetry, $\tau$, and the reflection symmetry, $\sigma_z$. 
Consequently, for nonzero wall-normal suction, the inversion symmetry $\sigma_{xyz}$ of PCF is broken.

The significance of symmetries for the dynamics of the flow is twofold. First, all flow fields which are invariant under the action of a symmetry contained in the equivariance group form a symmetry subspace within the system's state space that the dynamics is invariant under. That means, if an initial condition $\mathbf{u}$ is located in a symmetry subspace, its evolution under the governing equations remains in the same symmetry subspace. Second, if an equilibrium or travelling wave solution is invariant under a symmetry and thus located in a symmetry subspace parametric continuation will not change the symmetry of the solution but the entire continuation branch remains in the same symmetry subspace. Symmetries can only change if the branch is created or terminates in a symmetry-breaking bifurcation such as a pitchfork bifurcation.

If an equilibrium or travelling wave solution is \emph{not} invariant under a symmetry contained in the equivariance group of the system the action of that symmetry on the solution creates an additional solution. Two equilibria / travelling waves that are related by that symmetry have the same global integrated properties such as dissipation. Hence, symmetry-related invariant solution branches are represented by a single curve in bifurcation diagrams presenting a typical global integrated property as a function of a control parameter.
 
The invariant solutions presented in this study including equilibria and travelling waves are identified numerically using the Newton-Krylov-Hookstep root-finding tools contained in Channelflow 2.0 (www.channelflow.ch, \cite{Gibson2019}). To construct the required Krylov subspace, Channelflow employs successive time integrations of the Navier-Stokes equations to evaluate the action of Jacobian by finite differencing. 
For integrating the Navier-Stokes equations in time, a third-order accurate semi-implicit backward differentiation scheme is used together with a pseudo-spectral Fourier-Chebychev-Fourier discretisation. In the homogeneous streamwise, and spanwise directions, we dealiase the nonlinear term according to the 2/3 rule. The computational domain has a length of $L_x = 4\pi$, a width of $L_z=24\pi$, and a height of $2H = 2$ and is descretised by $N_x=32$, $N_y=35$, and $N_z=432$ collocation points. All of the solutions are localised in spanwise directions and thus independent of the width of the computational domain.


\section{Symmetry properties of snaking solutions in plane Couette flow without suction}\label{no_suction}

\begin{figure}
  \centerline{\includegraphics[width=0.9\textwidth]{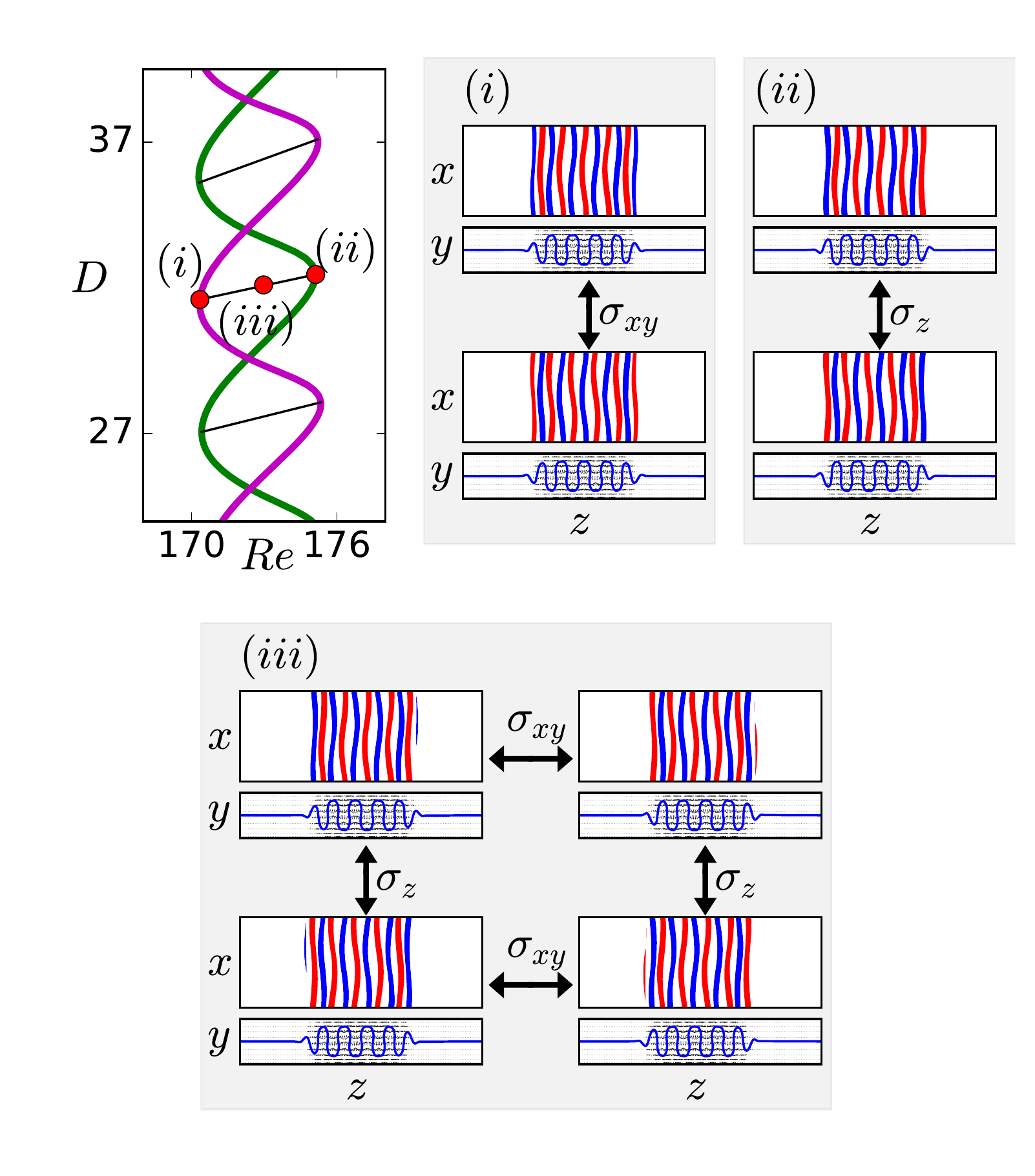}}
  \caption{(top left) A part of the snakes-and-ladders bifurcation structure of the localised invariant solutions in plane Couette flow in a box with a length of $4\pi$ and a width of $24\pi$. The symmetry-related solutions are visualised at the points indicated on the bifurcation diagram: The flow fields at point $(i)$ are two symmetry-related travelling waves; point $(ii)$ corresponds to two symmetry-related equilibrium solutions; and point $(iii)$ represents four symmetry-related rung states. All flow fields are visualised in terms of the midplane streamwise velocity (red/blue contours) and the streamwise averaged cross-flow velocity (vector plots). In addition, the contour line of zero average streamwise velocity is presented. The red/blue streaks indicate $\pm 0.2 U_w$. The entire computational domain is shown. Note the localisation of all solutions in spanwise direction.}
\label{fig:snaking_with_symms}
\end{figure}

\begin{figure}
  \centerline{\includegraphics[width=0.7\textwidth]{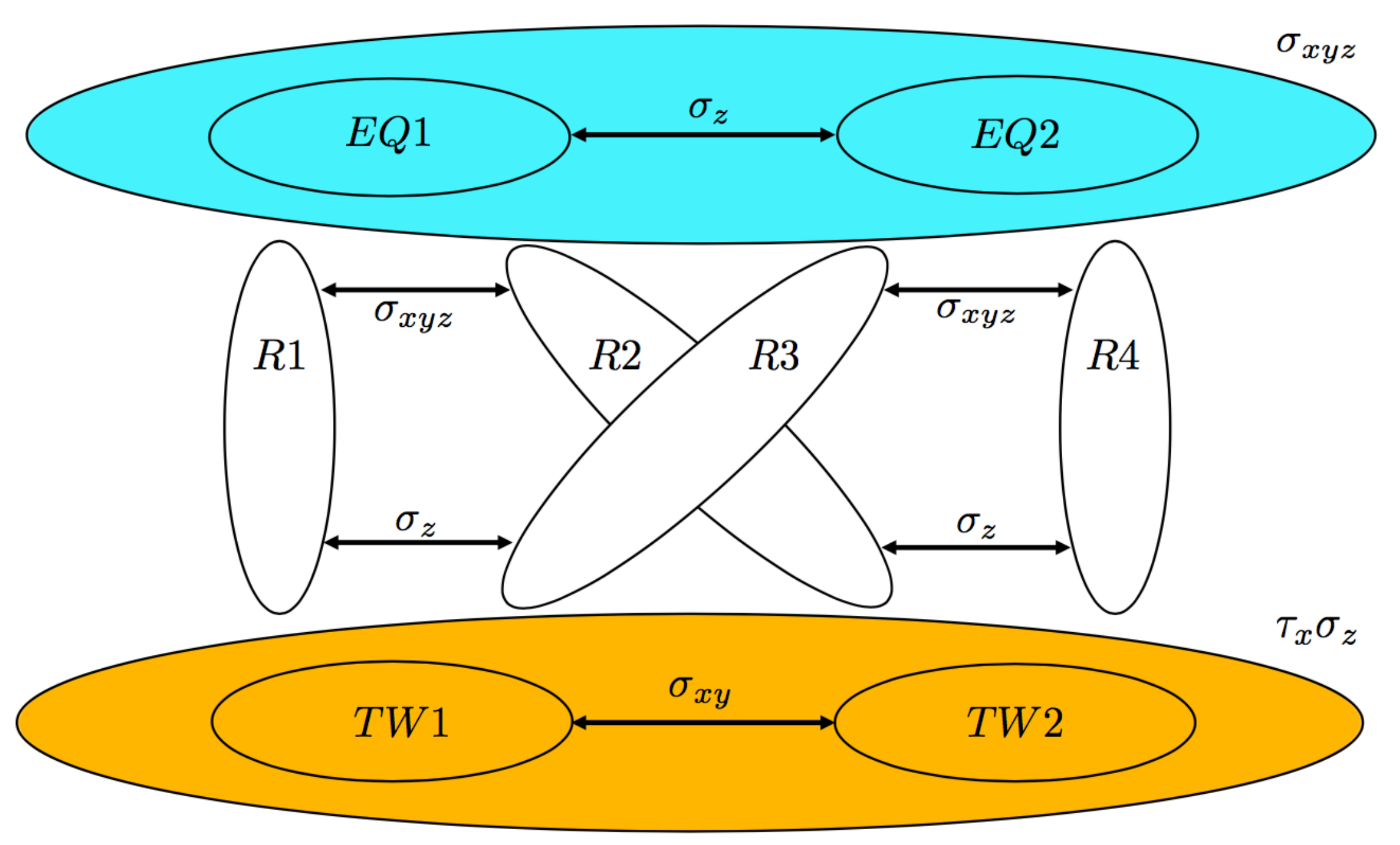}}
  \caption{Symmetry relations (arrows) between solution branches of plane Couette flow without suction. The pair of travelling waves $TW1/TW2$ is located within an invariant symmetry subspace (shaded orange ellipse). Likewise, the pair of equilibria $EQ1/EQ2$ shares a symmetry subspace (shaded cyan ellipse). All rungs $R1$, $R2$, $R3$ and $R4$ are non-symmetric. }
\label{fig:schematic_no_suction}
\end{figure}

The snaking solution found by \cite{Schneider2010} in plane Couette flow are shown in figure \ref{fig:snaking_with_symms} together with their bifurcation diagram. The solutions are computed for a periodic box with length $L_x=4\pi$ and show the snakes-and-ladders structure characteristic of homoclinic snaking. Two snaking curves winding upward in dissipation while oscillating in Reynolds number represent equilibrium (EQ) and travelling wave (TW) solutions, respectively. Each of the curves, both for EQs and TWs represents two symmetry-related solution branches. Pairs of equilibrium solutions and pairs of travelling waves are related by the rotational symmetry of PCF, $EQ1 = \sigma_{xy}EQ2$, and $TW1 = \sigma_{xy}TW2$. The operation of the discrete rotational symmetry leaves global solution measures such as the norm or the streamwise averaged energy dissipation unchanged. Consequently, both symmetry-related branches appear as a single curve in the dissipation $D$ versus $Re$ bifurcation diagram. In the snaking region between approximately $169 < Re < 177$ the dissipation of the snaking solutions increases. This is linked to the solution growing in spanwise direction while undergoing sequences of saddle-node bifurcations that create additional streaks at the solution fronts. The internal structure of the solutions remains unchanged. 

The equilibrium solutions are invariant under the inversion symmetry, $\sigma_{xyz}$ which is the product of the rotational symmetry, $\sigma_{xy}$, and the reflection symmetry, $\sigma_z$. A non-trivial invariant solution which is invariant under inversion is phase locked and can neither travel in the streamwise nor the spanwise directions so that the solution is an equilibrium and not a travelling wave. In contrast, the travelling wave solutions are invariant under a different symmetry, the shift-reflect symmetry, $\tau_x\sigma_z$ which locks only the $z$ phase of the solutions. As a result the travelling waves are free to travel in the streamwise direction. 
Since the equilibrium solutions are invariant under the inversion symmetry ($\mathbf{u}=\sigma_{xyz}\mathbf{u}$), the rotational symmetry relation between the two equilibria $EQ1=\sigma_{xy}EQ2$ reduces to $EQ1=\sigma_{xy}\sigma_{xyz}EQ2=\sigma_z EQ2$. Consequently, the rotational symmetry relation between the equilibrium solutions is equivalent to a mirror $z$-reflection symmetry relation between them. The two symmetry-related equilibria are thus mirror images with respect to the spanwise direction. For the travelling waves, their rotational symmetry relation implies that $TW1$ and $TW2$ travel in opposite directions but at equal phase speed.

In addition to the snaking branches, there are non-symmetric localised solutions forming rungs connecting the equilibrium branches to the travelling wave branches. Rungs are created in pitchfork bifurcations close to the saddle-node bifurcations along the snaking branches. There are four rung solution branches corresponding to each curve in the bifurcation diagram. They each connect one of the two equilibrium branches to one of the two travelling wave branches. Every pitchfork bifurcation on a specific equilibrium branch, creates two rung states which each connect this specific equilibrium branch to one of the two travelling waves. Since the rung states are non-symmetric invariant solutions which are created in symmetry-breaking pitchfork bifurcations off symmetric invariant solutions, the two simultaneously created rungs are related to each other by the symmetry of the symmetric solution branch which is broken. This is either the inversion symmetry of the equilibria $R1=\sigma_{xyz}R2$ and $R3=\sigma_{xyz}R4$ or the shift-reflect symmetry of the travelling waves. The shift-reflect symmetry relation between the two rung state solutions connecting to one travelling wave can be interpreted as only a $z$-reflection symmetry relation $\sigma_z$ between the rungs $R1 = \sigma_z R3$ and $R2 = \sigma_z R4$, because the shift is absorbed in the continuous translation symmetry for a solution that is not phase locked but free to travel in the streamwise direction. Hence at any Reynolds number, the four rung state solutions have the same dissipation, and the four branches of the rung states in the bifurcation diagram are represented by a single curve.
The symmetry relations between all branches of the snakes-and-ladders structure in PCF are schematically summarised in figure \ref{fig:schematic_no_suction} together with all relevant invariant symmetry subspaces. The two equilibria are located in the inversion symmetry subspace and they are related to each other by a $z$-reflection symmetry. The two travelling waves are in the subspace of the shift-reflect symmetry, and are related by the rotational symmetry, $\sigma_{xy}$. The rung state solutions are non-symmetric and live outside these two symmetry subspaces. 

\section{Modified snakes-and-ladders bifurcation structure  for non-vanishing suction}\label{modified_snaking}
Wall-normal suction breaks the inversion symmetry of PCF and leads to modifications of all discussed solution branches. In the following we first discuss modifications of the bifurcation diagram due to suction. Then, we present how solutions along the solution branches are modified.


\subsection{Effect of suction on the bifurcation diagram}\label{bifurcation_suction}

\begin{figure}
     \centering
    \subfloat{{\includegraphics[width=0.32\textwidth]{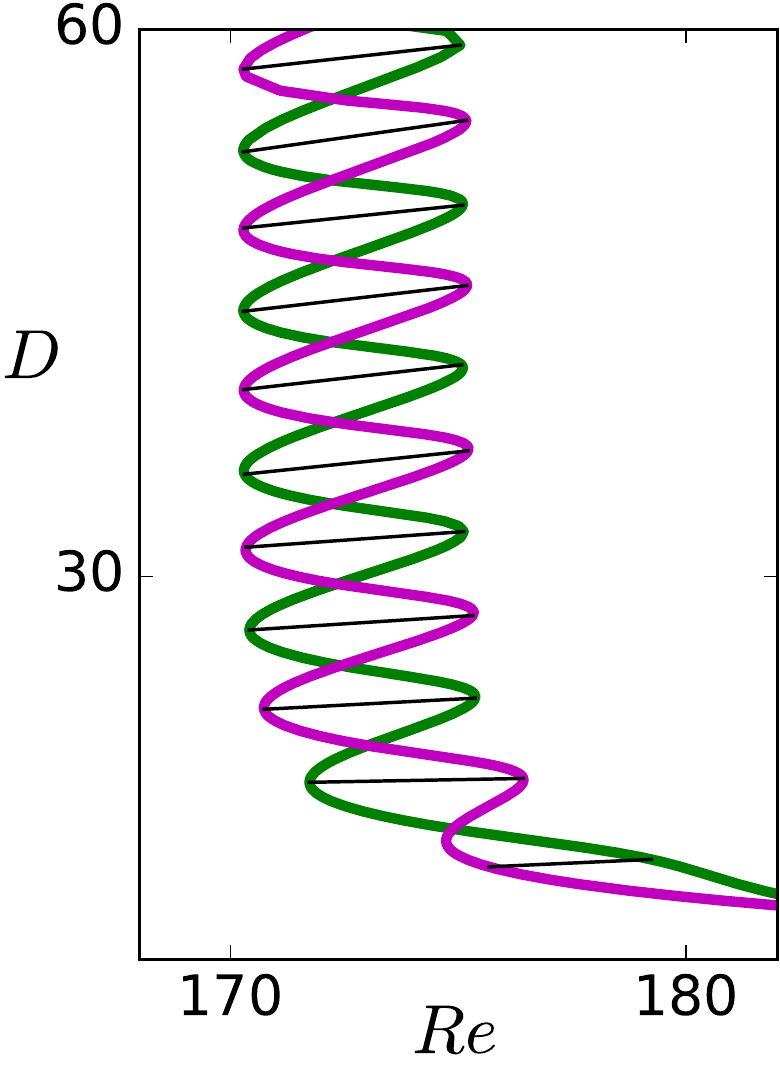} }}
    \subfloat{{\includegraphics[width=0.32\textwidth]{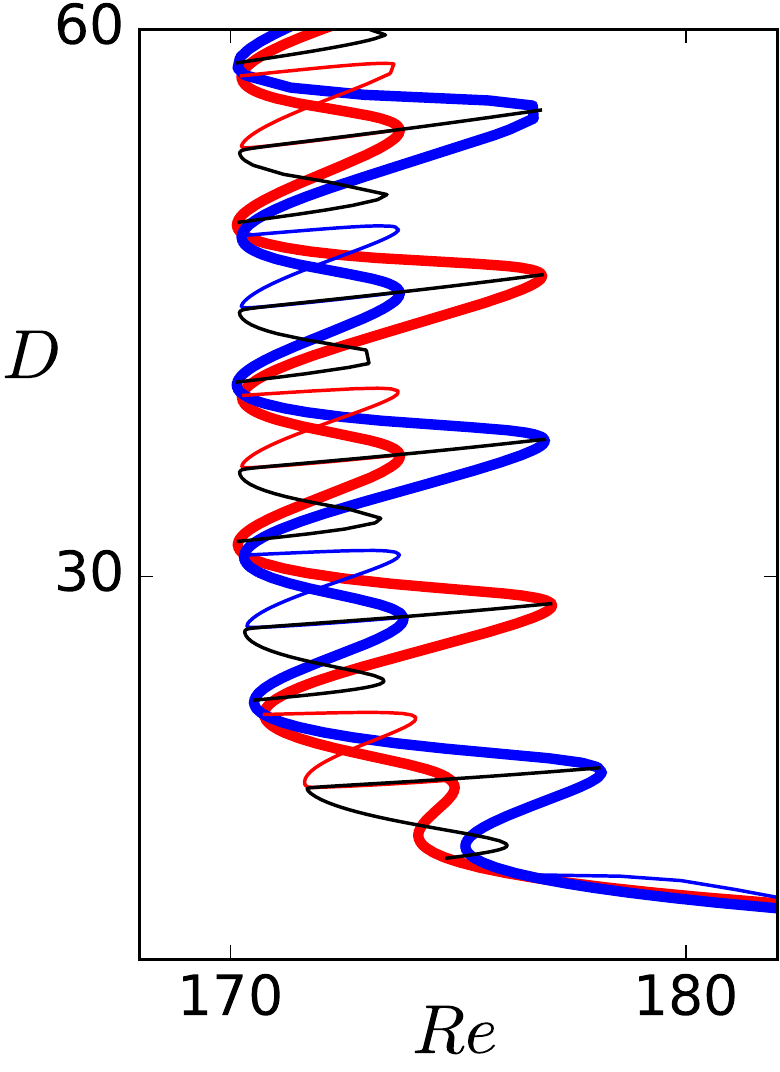} }}
    \subfloat{{\includegraphics[width=0.32\textwidth]{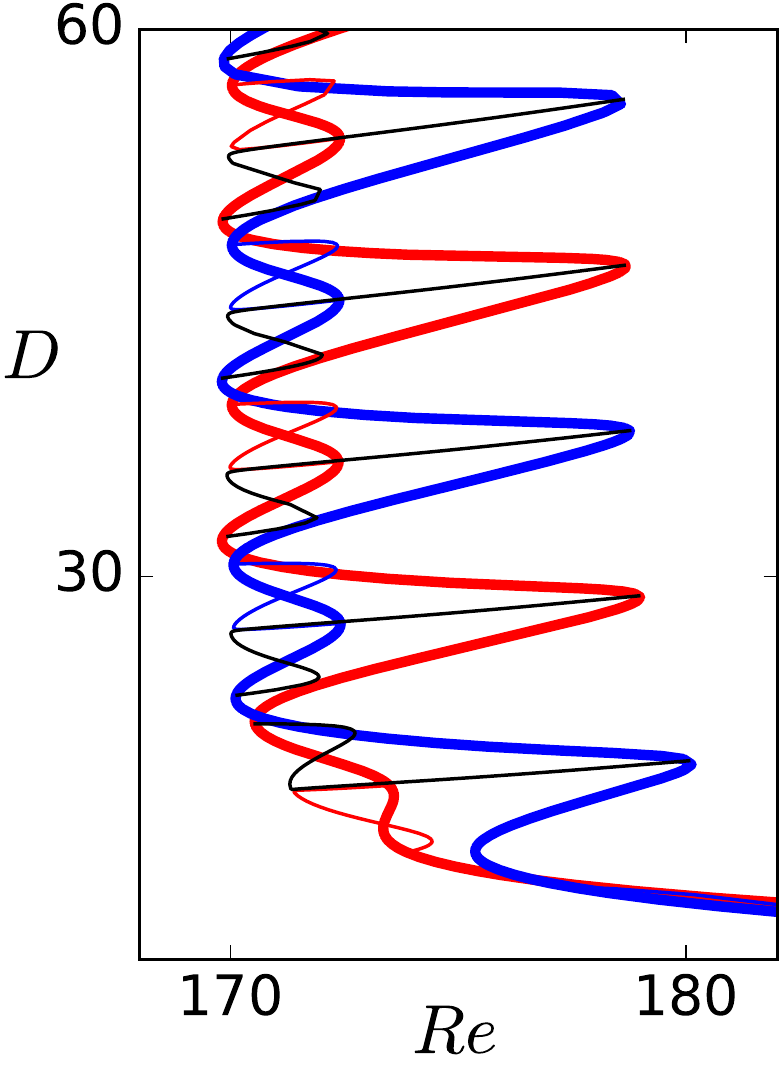} }}
    \caption{Modification of the snaking diagram with increasing wall-normal suction. The suction velocity is increased from $V_s = 0$ (left) to $V_s=10^{-4}$ (center) and $V_s=2\cdot 10^{-4}$ (right). Left panel: Travelling wave (magenta) and equilibrium (green) snaking branches are shown together with rungs (black). 
    Center and right panels: The travelling wave branch splits into $TW1$ (thick red) and $TW2$ (thick blue). The snaking equilibrium branch has broken into disconnected segments and together with remainders of rungs forms \emph{returning states} $RS$ and \emph{connecting states} $CS$. Two types of $RS$ (thin red / blue) connect to $TW1$ / $TW2$, respectively. The $CS$ states (thin black) connect $TW1$ and $TW2$.}
    \label{fig:snaking_with_suction}
\end{figure}

\begin{figure}
    \centering
    ($a$)
    \subfloat{{\includegraphics[width=0.45\textwidth]{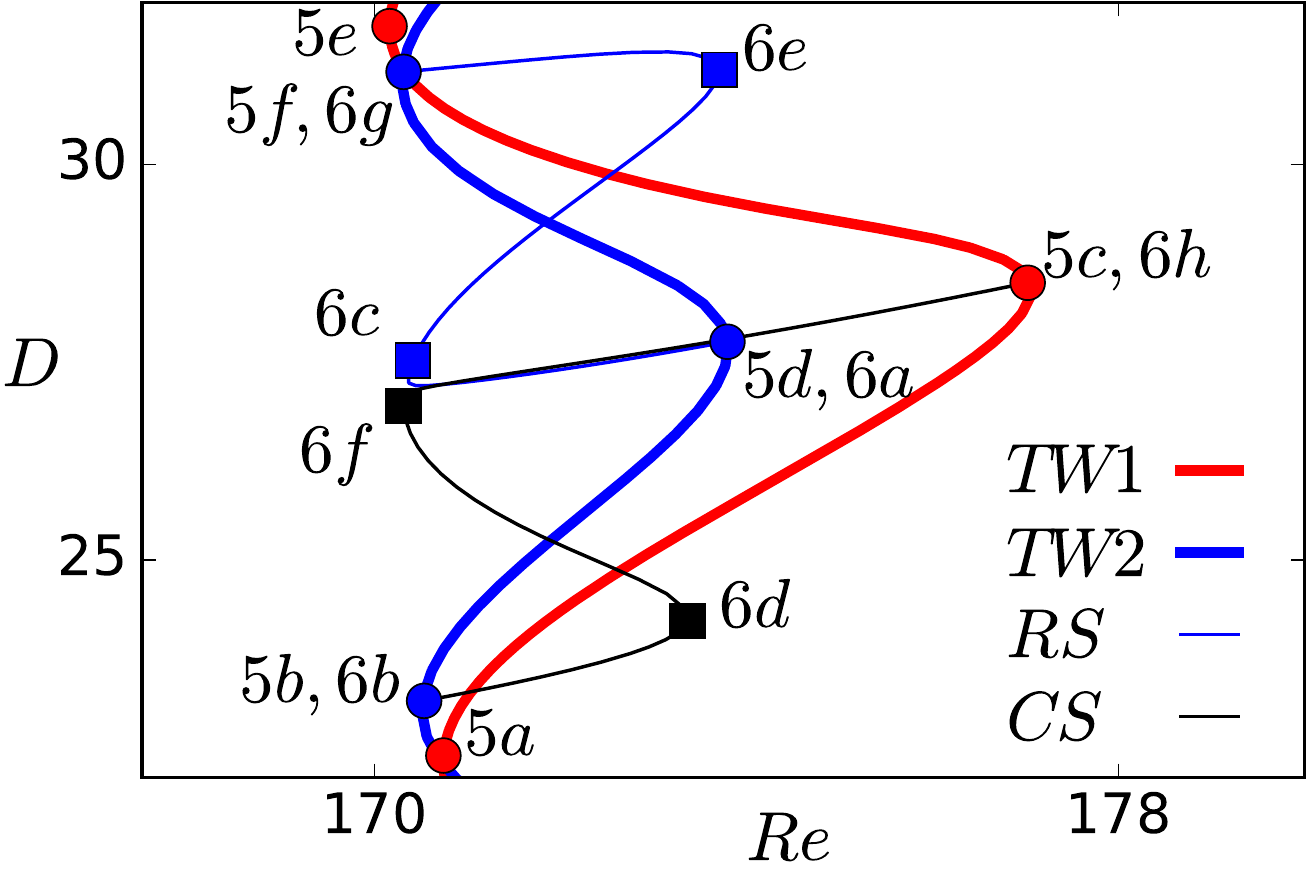} }}
    ($b$)
    \subfloat{{\includegraphics[width=0.45\textwidth]{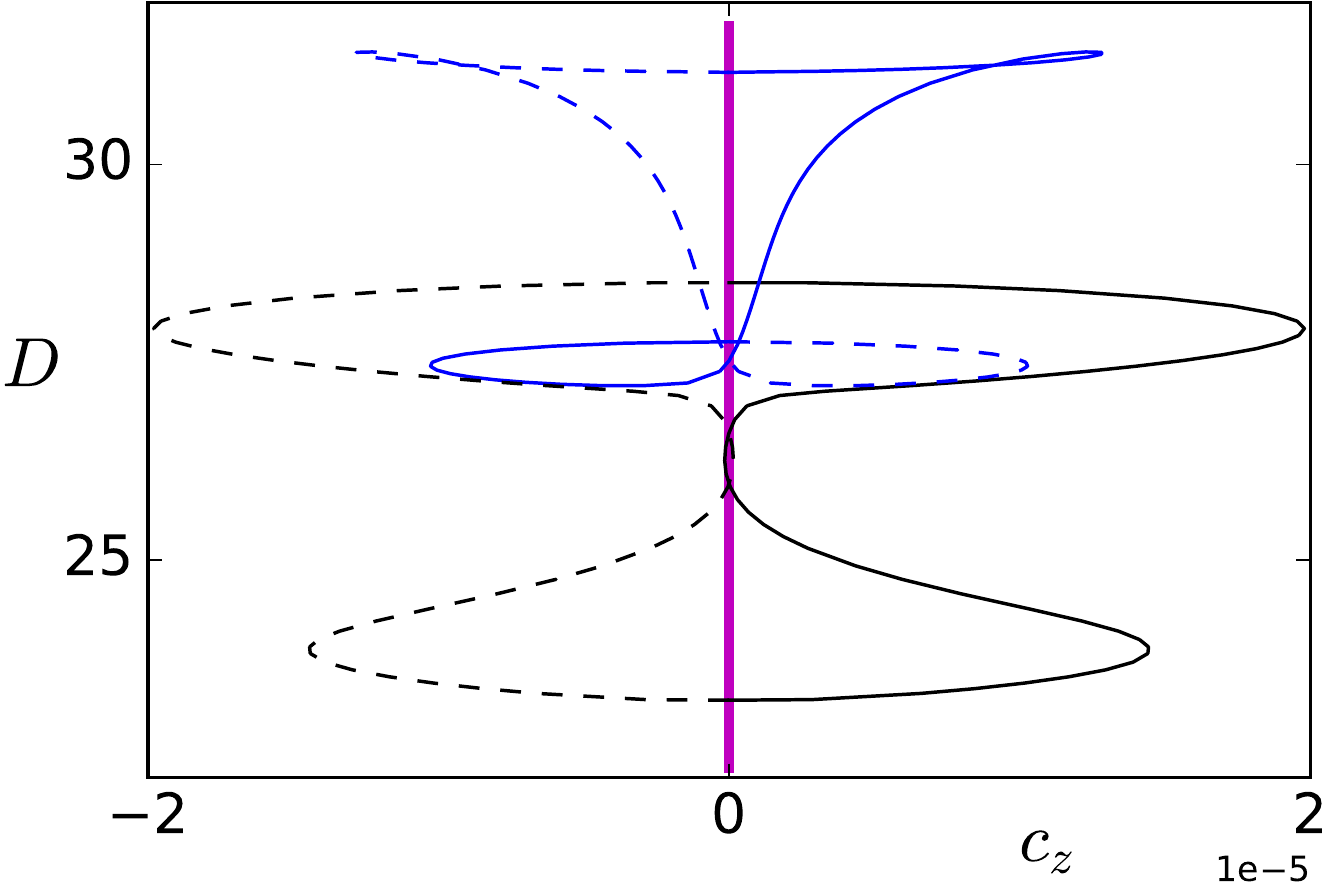} }}
    \caption{ $(a)$ One oscillation of the two travelling waves. The flow fields at the indicated points are visualised in figures \ref{fig:TW_fields} and \ref{fig:RS_CS_fields}. The colour coding is the same as in figure \ref{fig:snaking_with_suction}. $(b)$ Variation of the spanwise wave speed illustrating the two symmetry-related $RS$ (solid and dashed blue) and $CS$ (solid and dashed black) branches.}
    \label{fig:one_course}
\end{figure}

Figure \ref{fig:snaking_with_suction} shows how the snakes-and-ladders bifurcation structure is modified when wall-normal suction is applied. The two travelling wave branches undergoing snaking are symmetry-related at zero suction and thus appear as a single curve. For non-zero suction, the symmetry relating both branches is not contained in the equivariance group. Consequently, the symmetry relation vanishes and both travelling wave branches split. 
The two travelling wave branches remain continuous and undergo snaking but the critical $Re$ at which saddle-node bifurcations occur alternates. Following the terminology of \cite{Knobloch2015}, we refer to the piece of a snaking curve between two forward saddle-node bifurcations including a backward saddle-node bifurcation as an \textit{oscillation} in the snaking branch. In the snaking region, both travelling wave branches are composed of two alternating oscillations with different spans in Reynolds number, a narrow oscillation and a wide oscillation. Where travelling wave branch $TW1$ undergoes a narrow oscillation the other one $TW2$ undergoes a wide oscillation. For the next pair of oscillations the situation reverses with, $TW1$ undergoing a wide, and $TW2$ a narrow oscillation. 

Equilibrium branches that undergo snaking in the zero-suction case are invariant under the inversion symmetry, which is broken by suction. As a result, for non-zero suction, the continuous equilibrium solution branches vanish. Instead the branch breaks into disconnected segments. 
The pitchfork bifurcations that -- for the zero suction case -- create rung states bifurcating from the equilibrium branches, are broken by the wall-normal suction. At non-zero suction the disconnected remainders of the equilibrium branches together with remainders of rung states form new branches that connect to travelling wave branches. These branches fall into two groups.
Some branches emerge in a pitchfork bifurcation  on one of the travelling wave branches and terminate on the same travelling wave branch in another pitchfork bifurcation. Hereafter, these branches are called \textit{returning state, $RS$}. There are two $RS$ branches which are related by a $z$-reflection symmetry, $RS1 = \sigma_z RS2$. 
All other non-symmetric branches connect $TW1$ to $TW2$. These branches are, hereafter, referred to as \textit{connecting state, $CS$}. In the continuation diagram, each $CS$ curve represents two symmetry-related branches, $CS1 = \sigma_z CS2$, which bifurcate from the travelling wave branches in pitchfork bifurcations.

The $RS$ branches as well as the $CS$ branches form closed bifurcation loops, as every branch has a symmetry-related counterpart starting and terminating in the same pitchfork bifurcation on $TW$ branches. In figure \ref{fig:one_course}($a$) a part of the bifurcation diagram enlarging one oscillation of the travelling waves is shown. The loops can be directly observed in figure \ref{fig:one_course}($b$) where solutions are shown in terms of dissipation versus spanwise wave speed. The spanwise wave speed differentiates between both $z$-reflection symmetry-related branches. The spanwise wave speed of the two symmetry-related branches has the same magnitude but opposite sign. Due to their shift-reflect symmetry, the spanwise wave speed of the travelling wave branches is zero.


\subsection{Evolution of the flow fields on the solution branches}\label{sec:flowfields_with_Vs}

\begin{figure}
    \centering
    ($a$)
    \subfloat{{\includegraphics[width=0.45\textwidth]{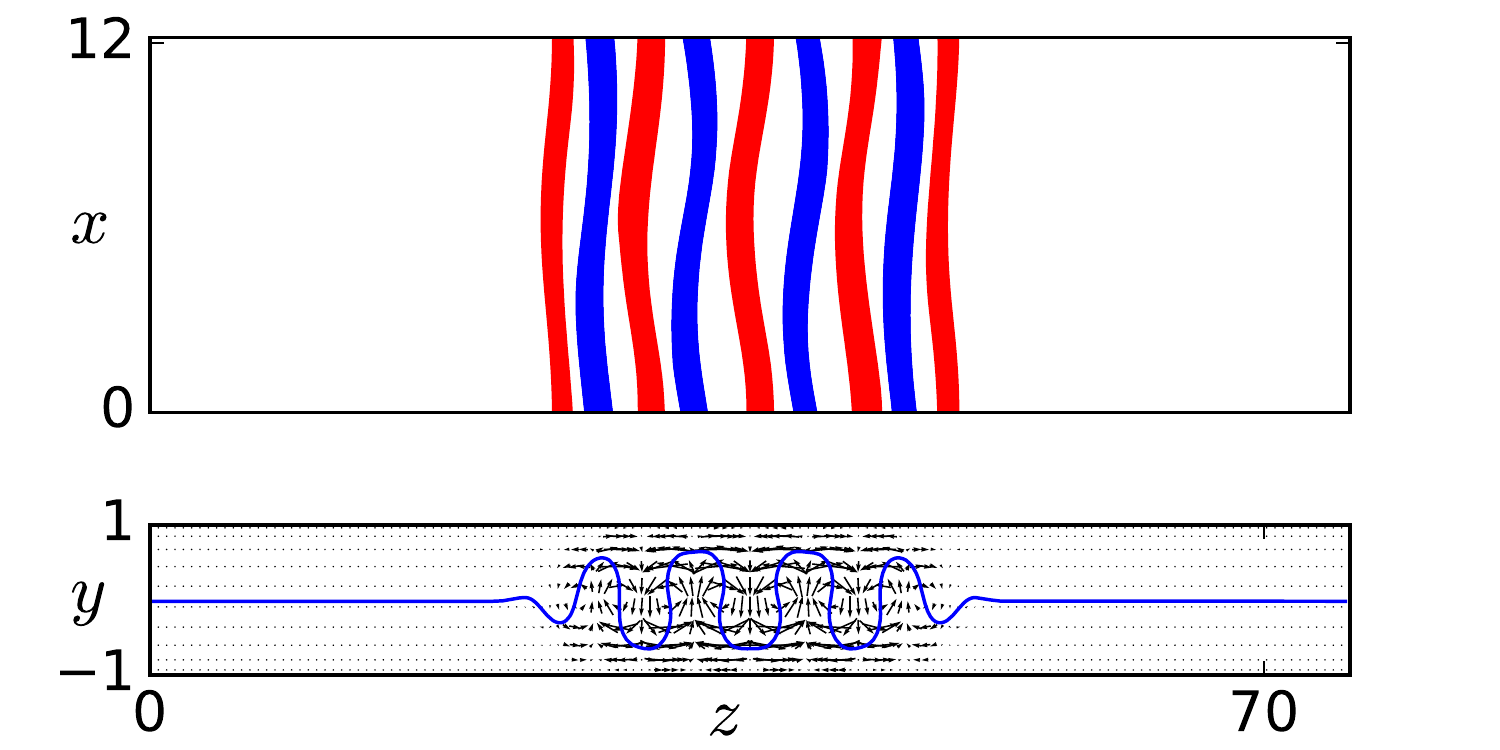} }}
    ($b$)
    \subfloat{{\includegraphics[width=0.45\textwidth]{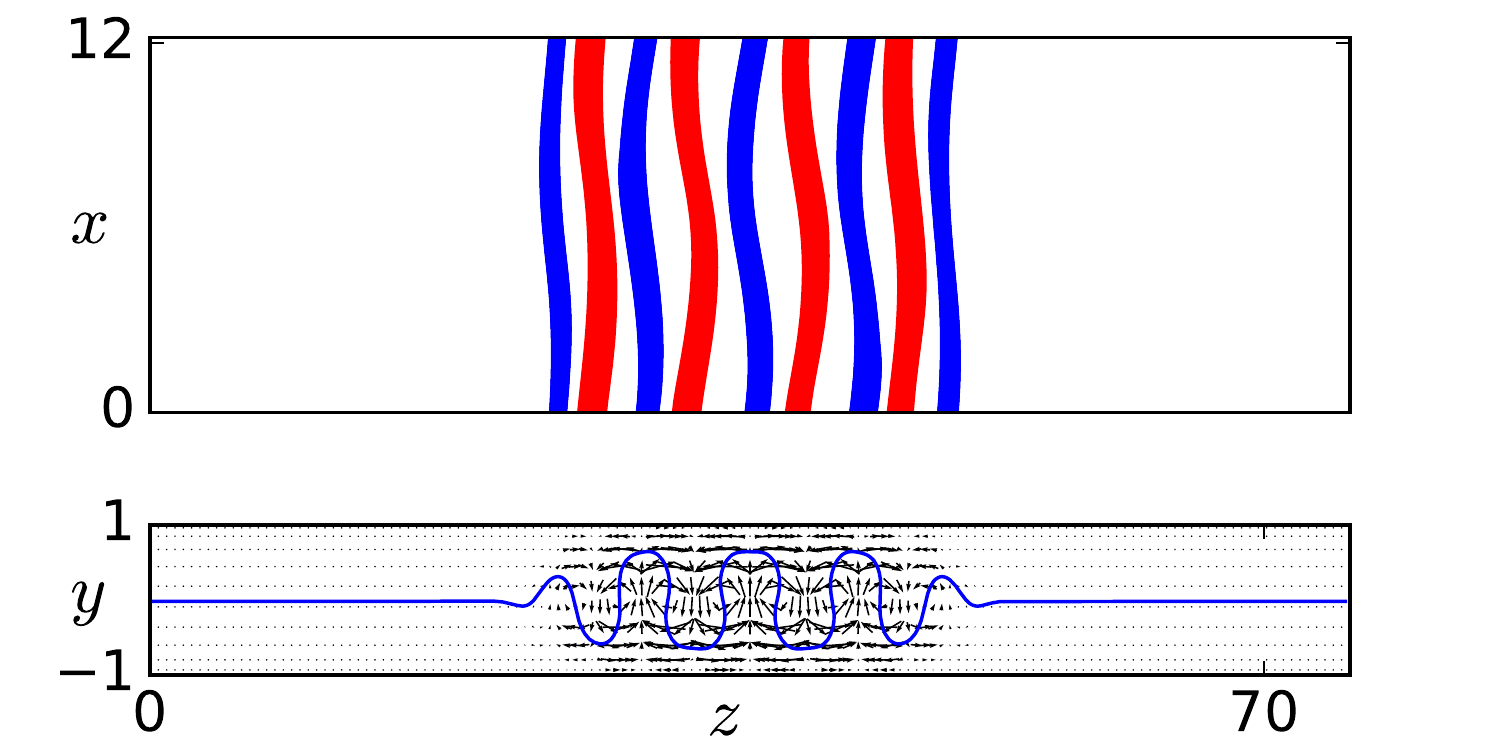} }}\\
    ($c$)
    \subfloat{{\includegraphics[width=0.45\textwidth]{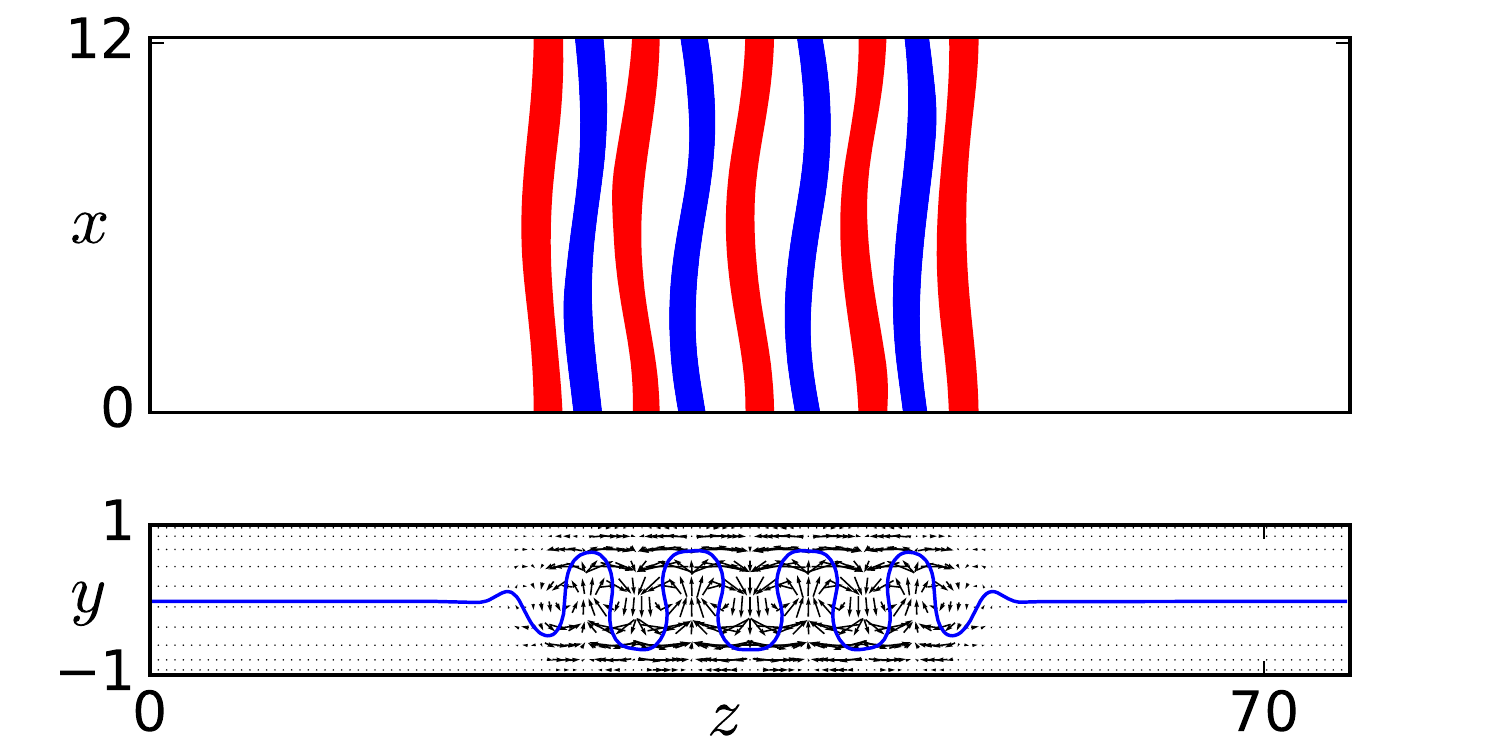} }}
    ($d$)
    \subfloat{{\includegraphics[width=0.45\textwidth]{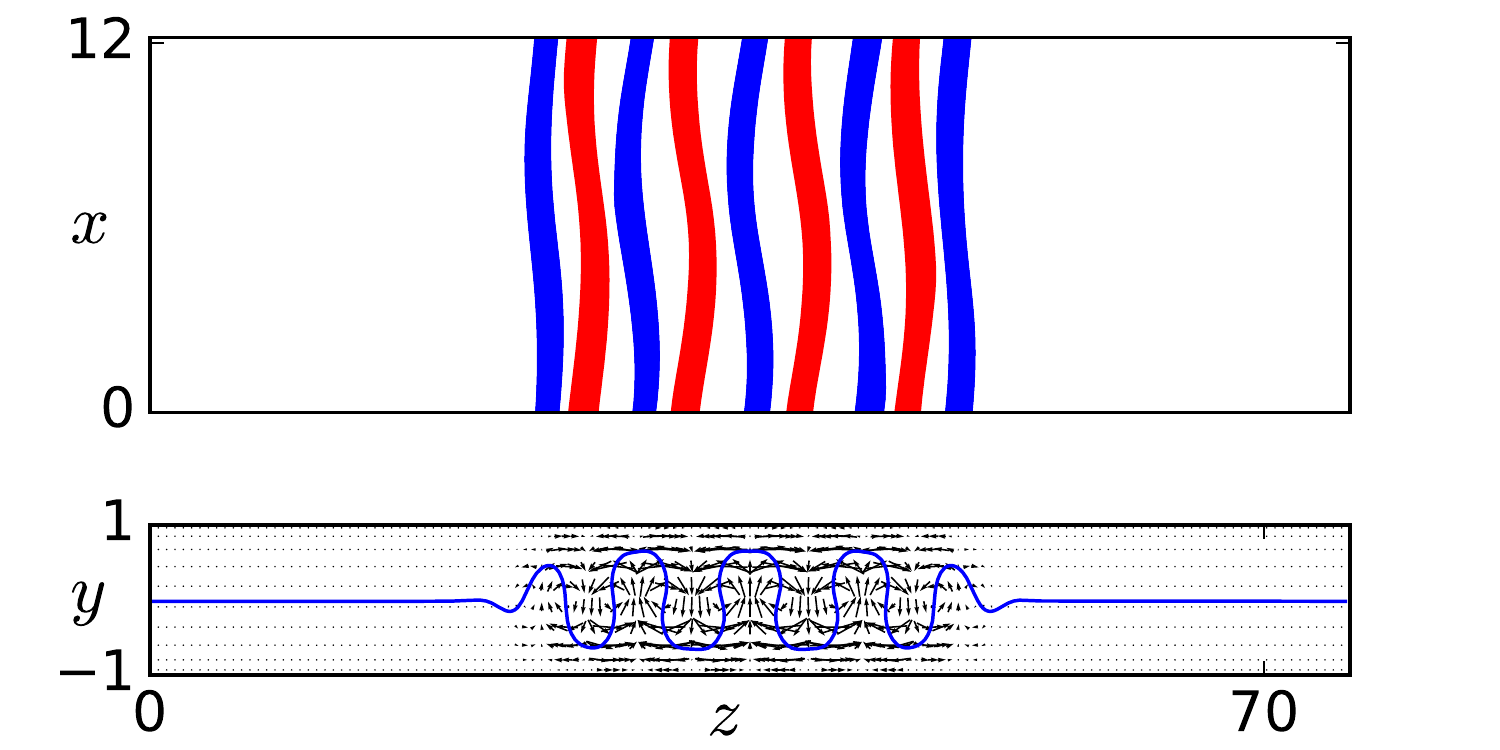} }}\\
    ($e$)
    \subfloat{{\includegraphics[width=0.45\textwidth]{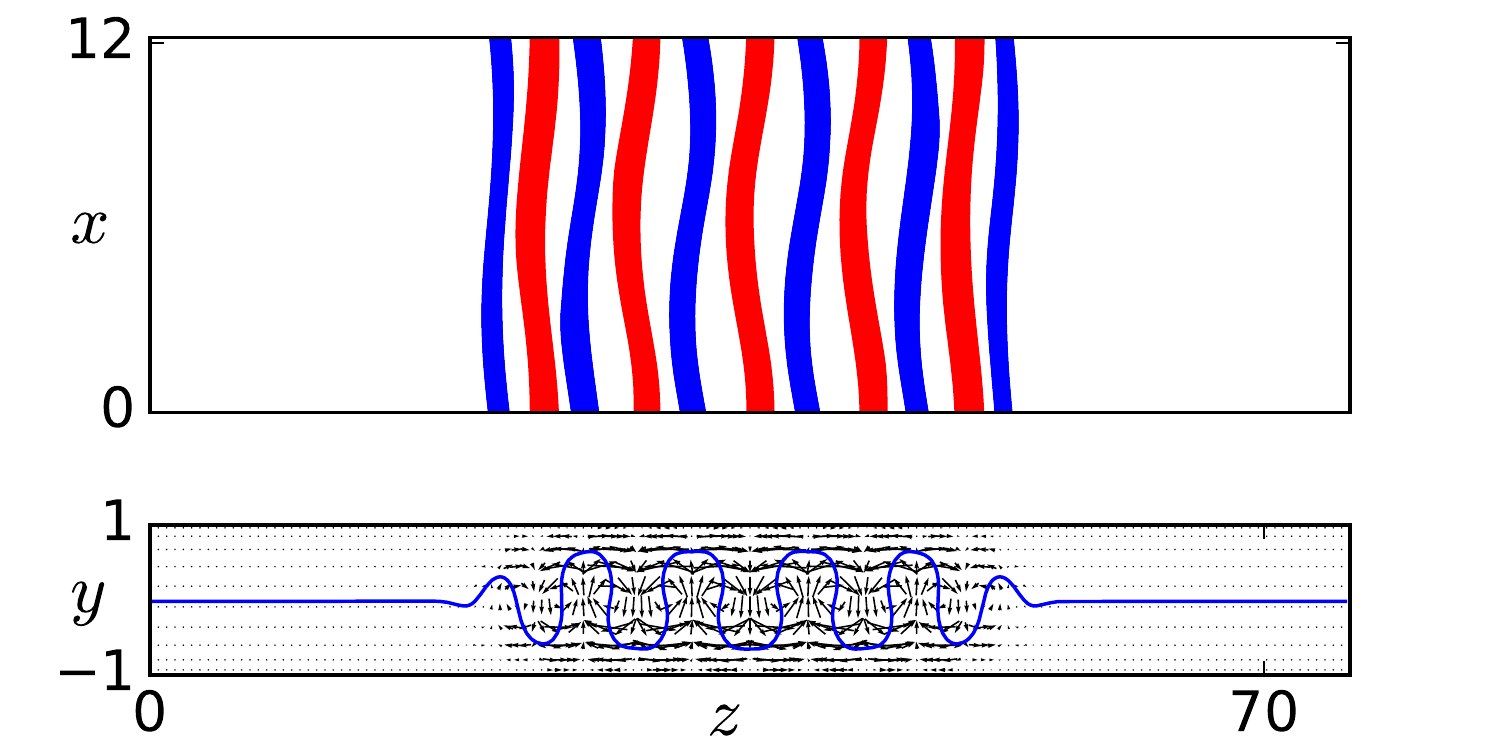} }}
    ($f$)
    \subfloat{{\includegraphics[width=0.45\textwidth]{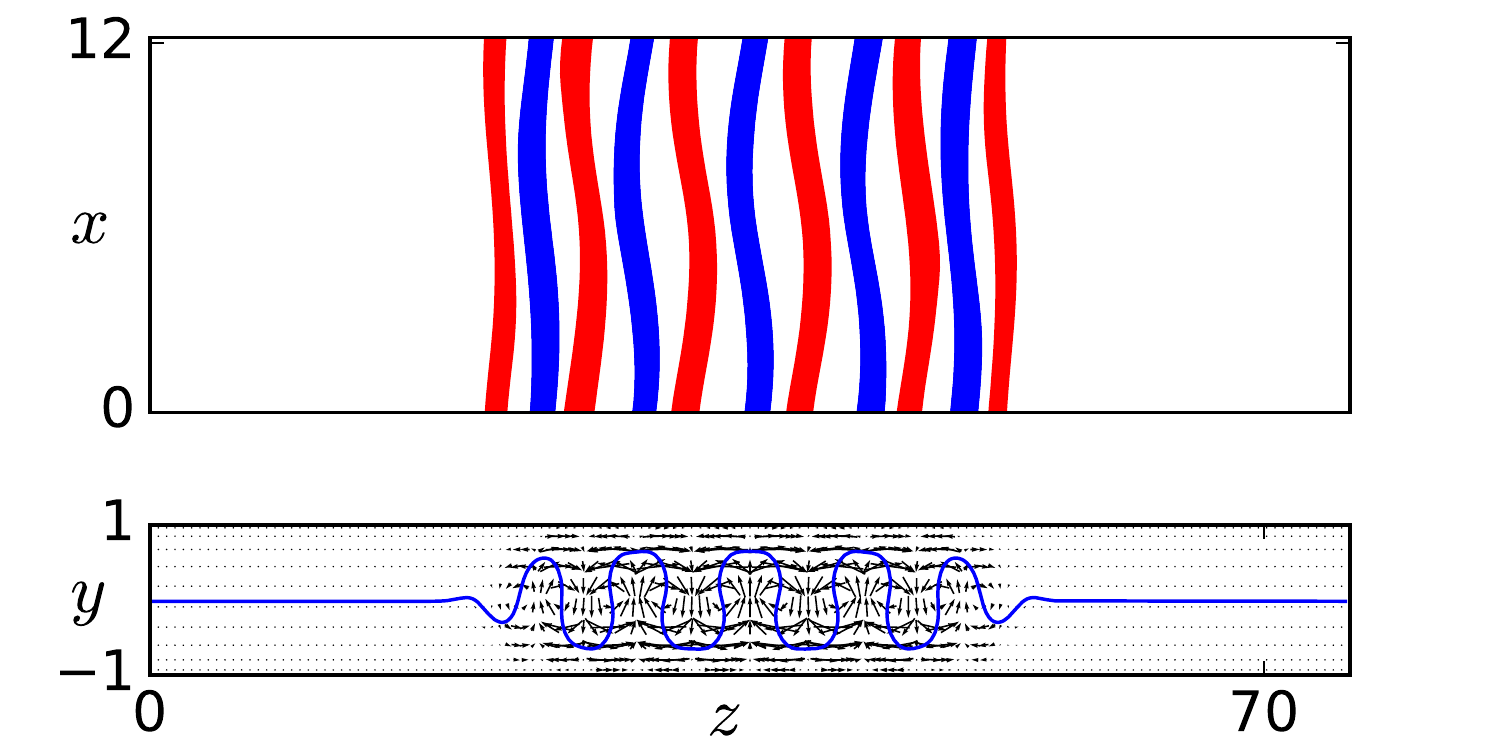} }}
    \caption{Flow fields at points indicated in figure \ref{fig:one_course}($a$) on the travelling wave branches. In the left panels the flow fields of $TW1$, and in the right panels the flow fields of $TW2$ are shown. The visualisation of the flow fields are the same as the visualisations in figure \ref{fig:snaking_with_symms}.}
    \label{fig:TW_fields}
\end{figure}

\begin{figure}
    \centering
    ($a$)
    \subfloat{{\includegraphics[width=0.45\textwidth]{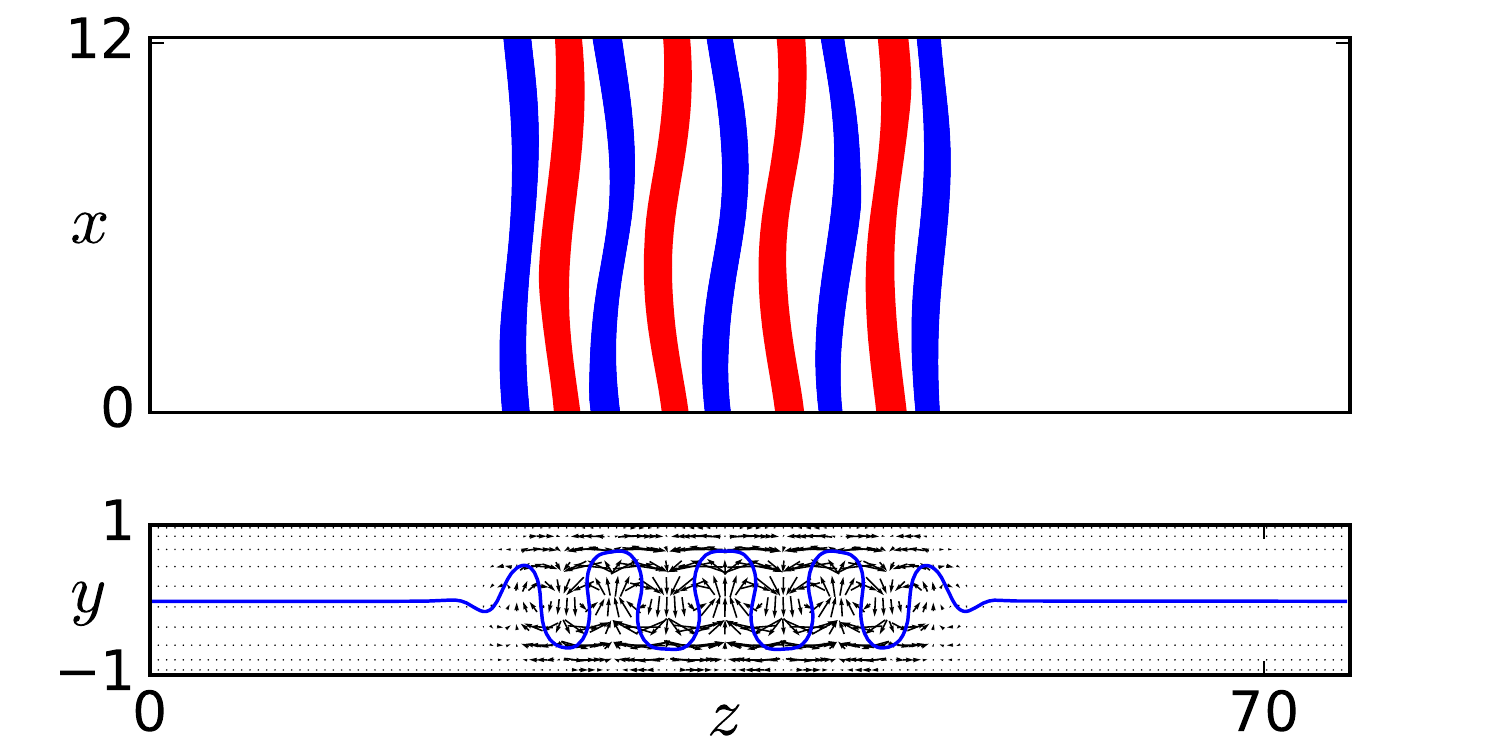} }}
    ($b$)
    \subfloat{{\includegraphics[width=0.45\textwidth]{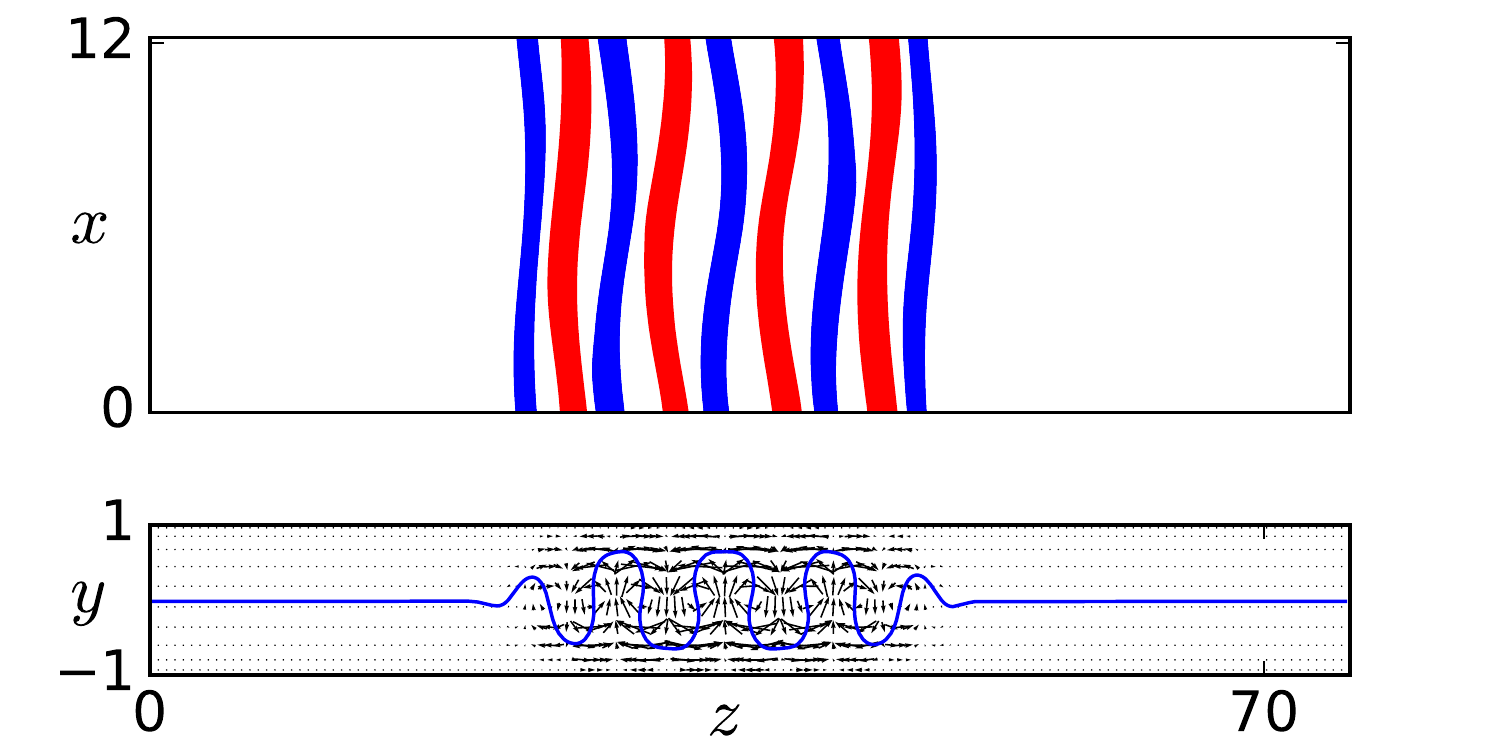} }}\\
    ($c$)
    \subfloat{{\includegraphics[width=0.45\textwidth]{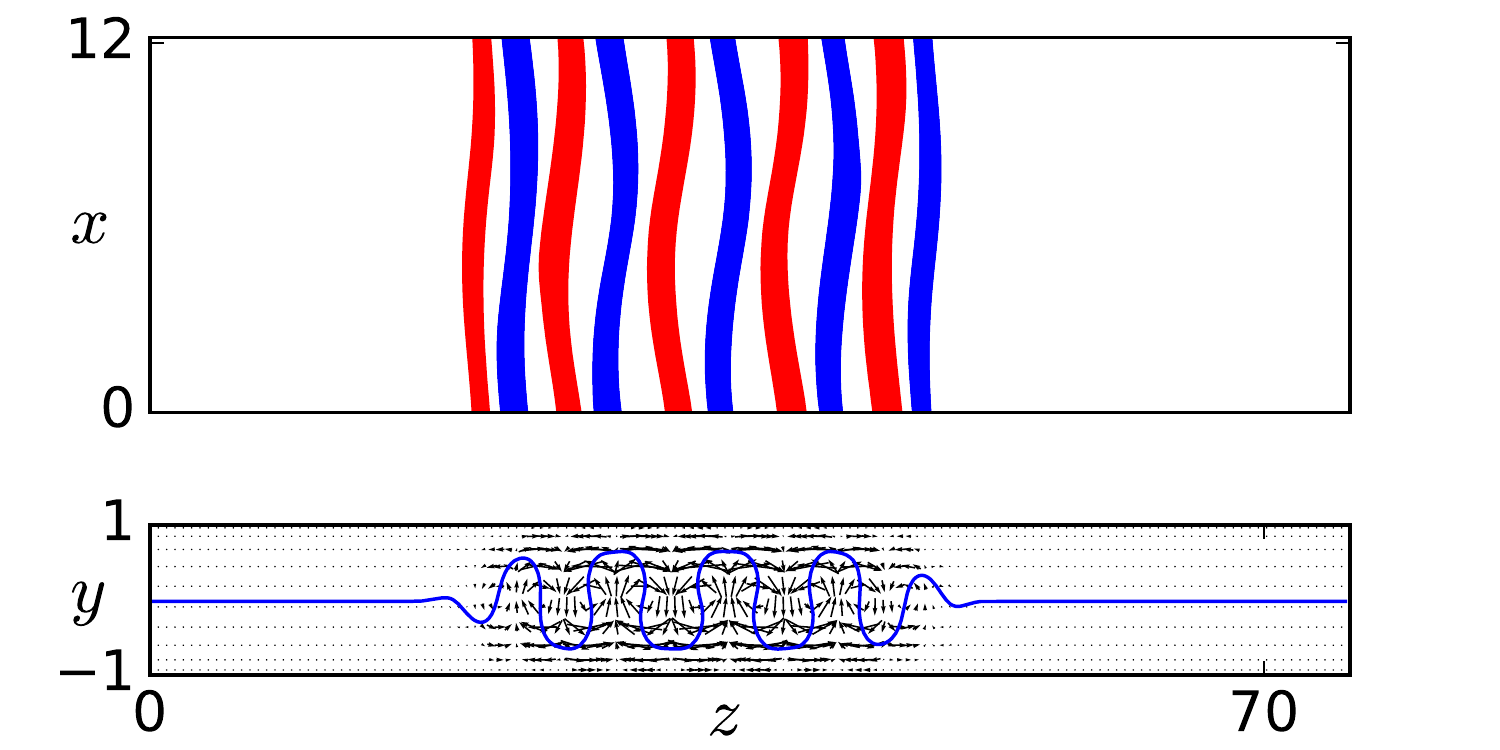} }}
    ($d$)
    \subfloat{{\includegraphics[width=0.45\textwidth]{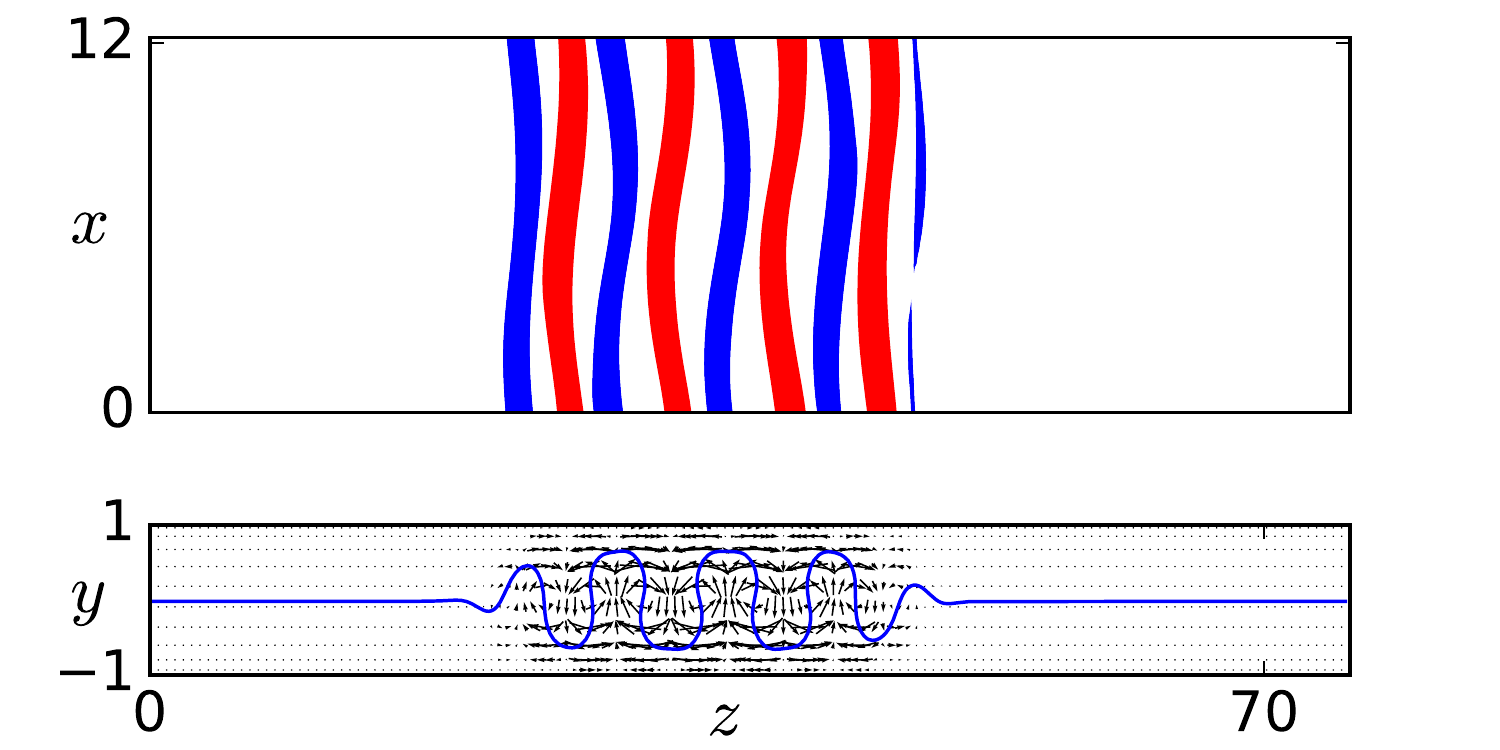} }}\\
    ($e$)
    \subfloat{{\includegraphics[width=0.45\textwidth]{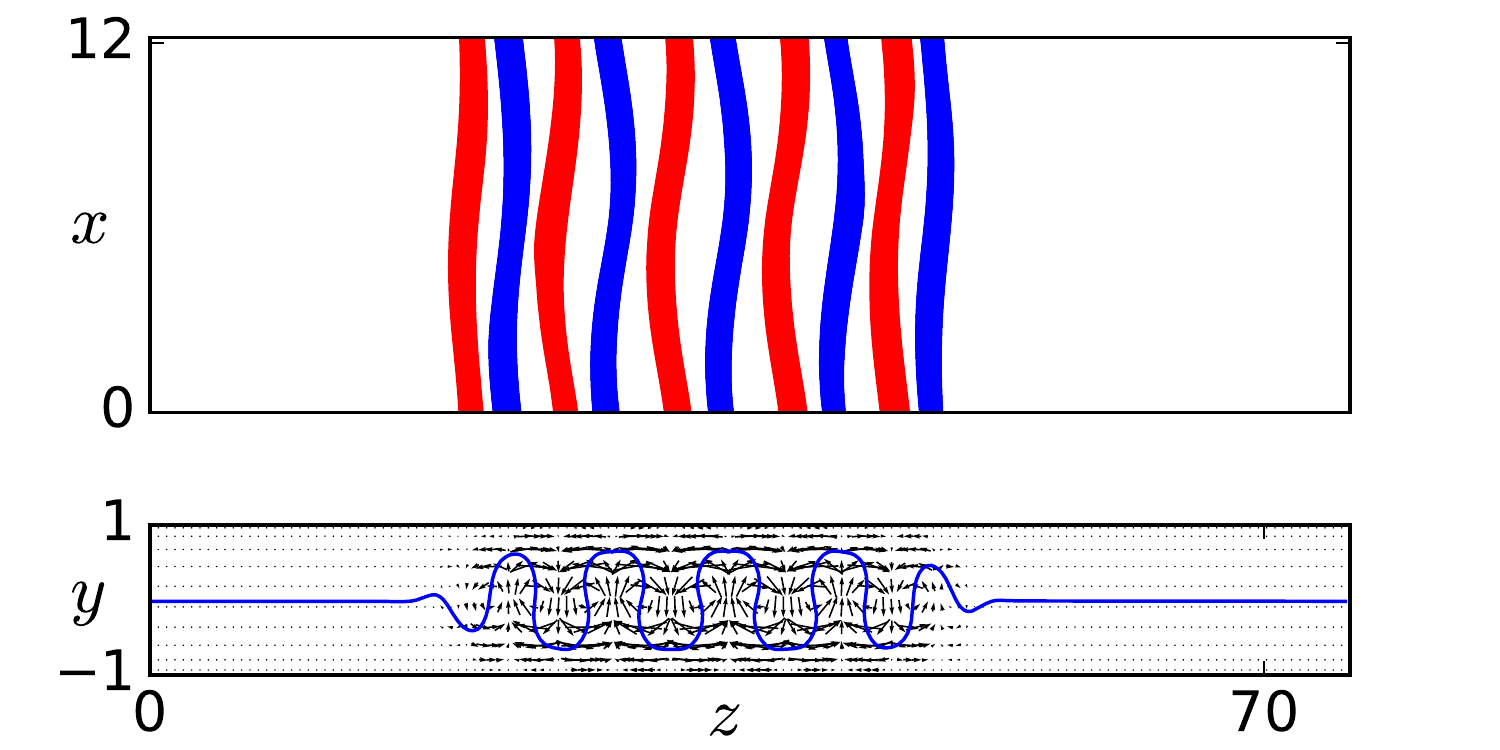} }}
    ($f$)
    \subfloat{{\includegraphics[width=0.45\textwidth]{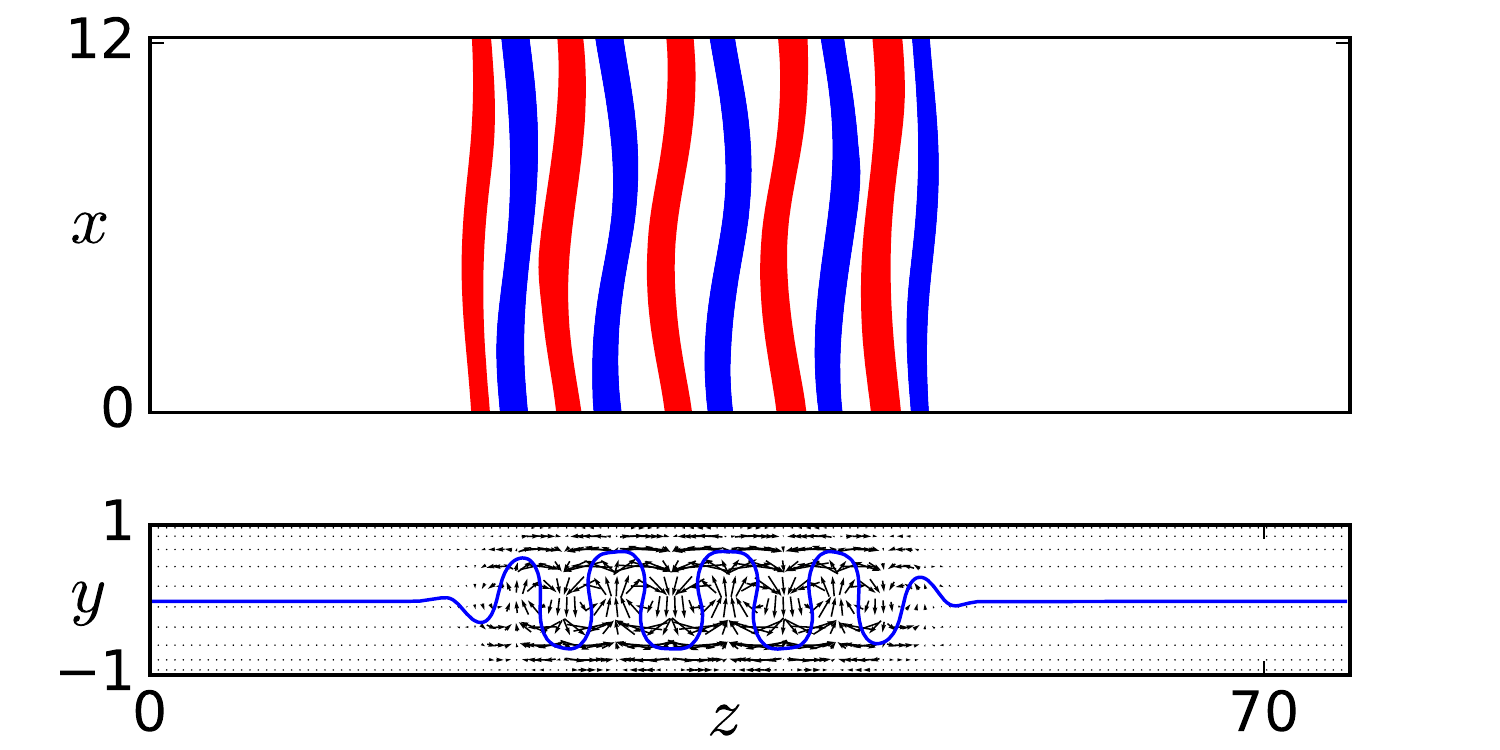} }}\\
    ($g$)
    \subfloat{{\includegraphics[width=0.45\textwidth]{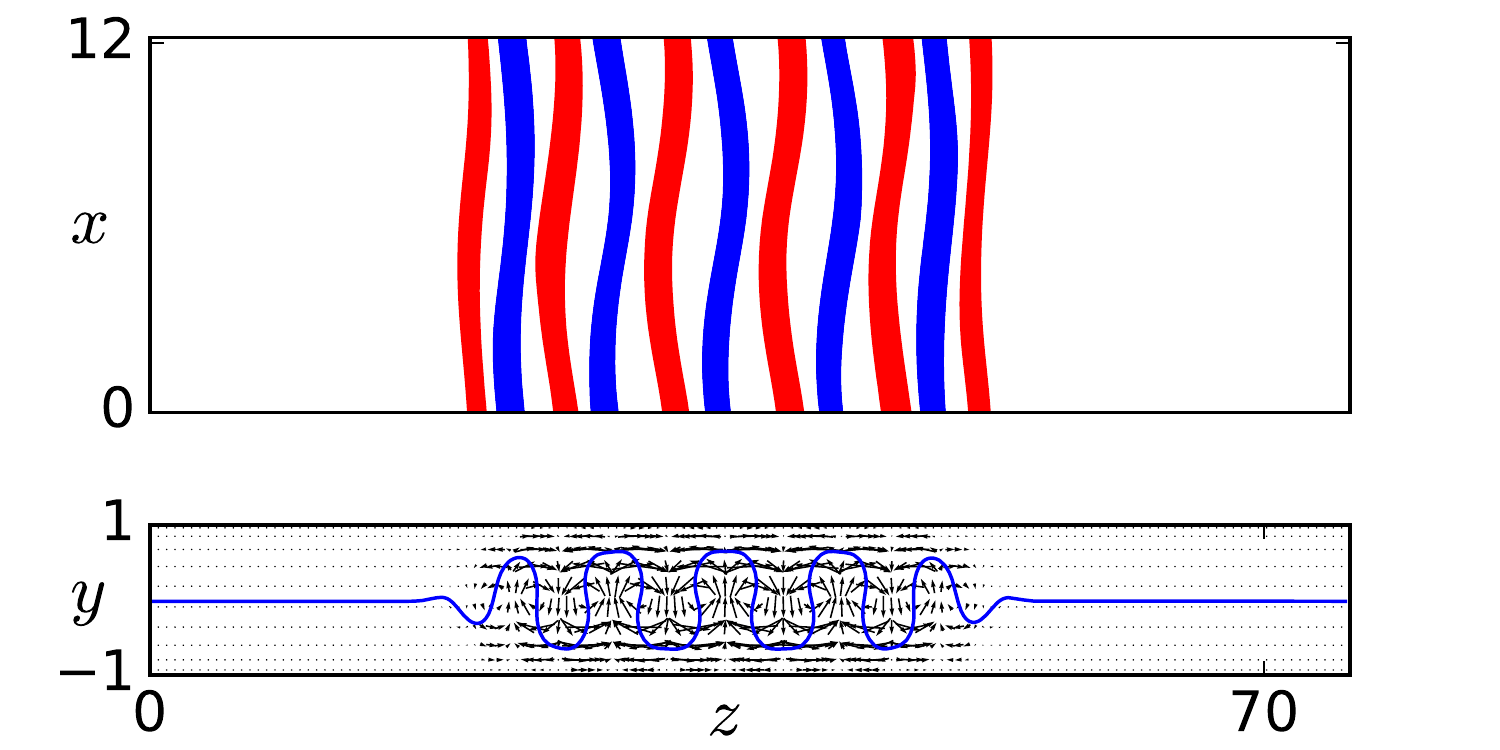} }}
    ($h$)
    \subfloat{{\includegraphics[width=0.45\textwidth]{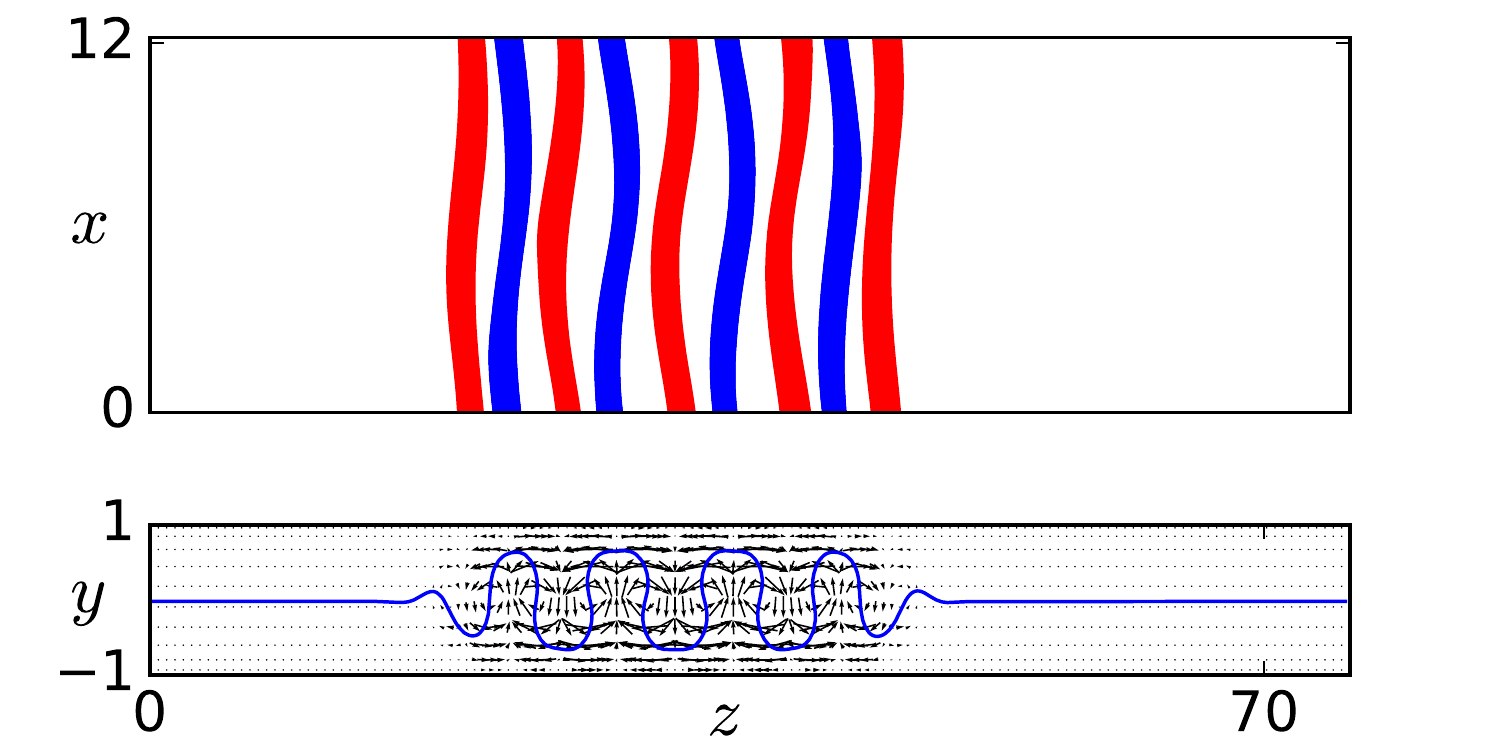} }}
    \caption{Flow fields of the returning state branch $RS$ (left) and the connecting state  branch $CS$ (right) at the points indicated in figure \ref{fig:one_course}($a$). The panel labels indicate the points in the continuation diagram in figure \ref{fig:one_course}($a$). The visualisation of the flow fields are the same as the visualisations in figure \ref{fig:snaking_with_symms}.}
    \label{fig:RS_CS_fields}
\end{figure}

\begin{figure}
    \centering
    \subfloat{{\includegraphics[width=0.45\textwidth]{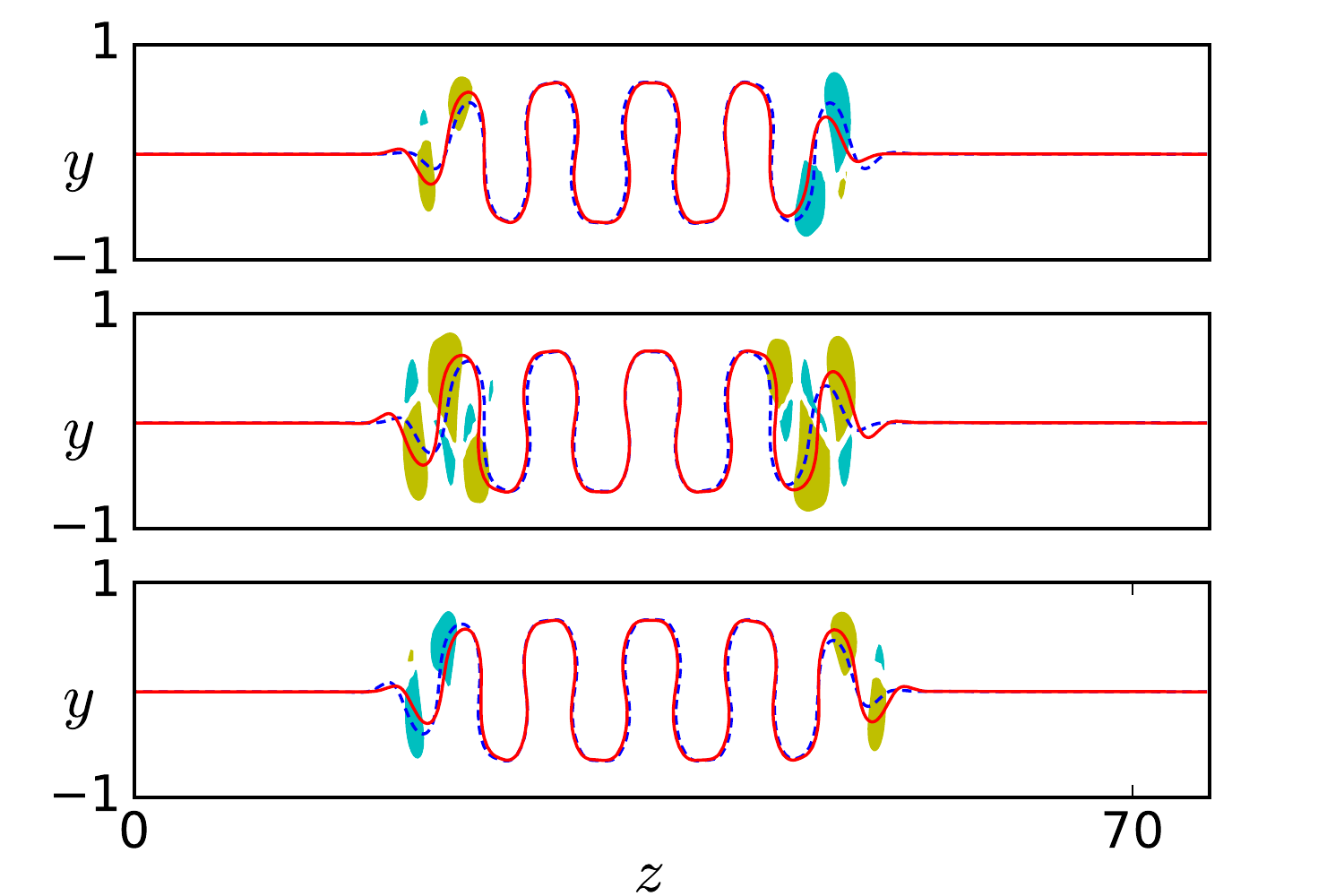} }}
    \subfloat{{\includegraphics[width=0.45\textwidth]{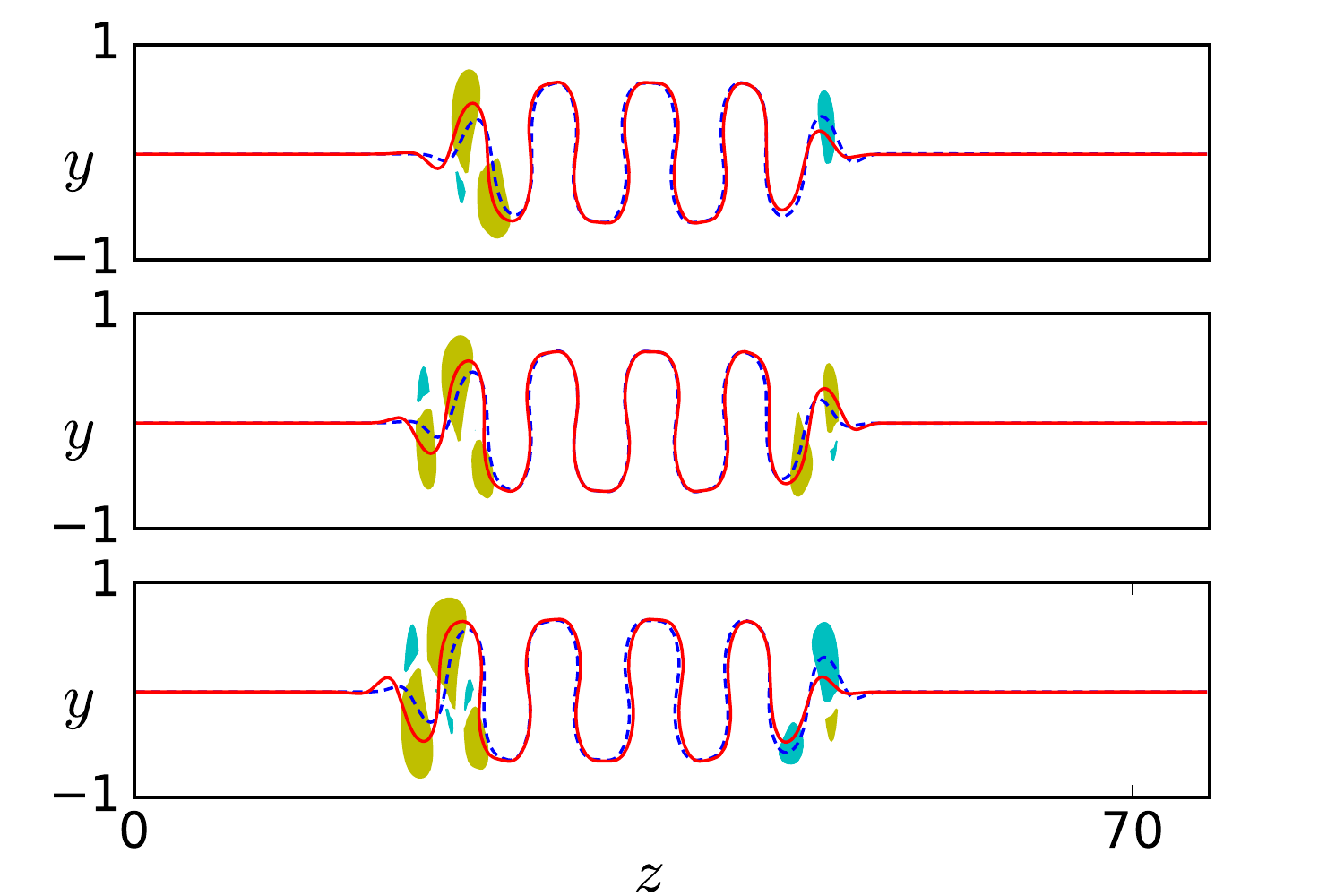} }}
    \caption{The variations of flow field along the $RS$ (left) and $CS$ branch (right): Overlay of zero-contour lines of average streamwise velocity for pairs of flow fields shown in figure \ref{fig:RS_CS_fields} as dashed blue / solid red line. (6a)/(6c) (top left), (6c)/(6e) (middle left), (6e)/(6g) (bottom left); (6b)/(6d) (top right), (6d)/(6f) (middle right), (6f)/(6h) (bottom right). Differences are visible close to the fronts. Colours indicate contours of the amplitude difference of the average streamwise velocity $|\langle u_2 \rangle_x| - |\langle u_1 \rangle_x|$ with yellow/cyan corresponding to $\pm 0.13 U_w$. Along the $RS$ branch (left), the envelope of the localised solutions shifts to the left, grows symmetrically and shifts to the right. After the sequence, the contour line remains centered at a maximum. Along the $CS$ branch the structure shifts left, grows symmetrically and shifts left again. As a result, the contour line in the center of the periodic pattern turns from a maximum into a minimum.}
    \label{fig:moving_frame}
\end{figure}

Each saddle-node bifurcation along the snaking branches is associated with spanwise growth of the solution as evidenced by the increasing value of dissipation. The neutral eigenmode associated with each saddle-node is localised at the fronts of the localised solution where the spatially periodic internal pattern connects to unpatterned laminar flow \citep{Schneider2010}. When the snaking branch undergoes a saddle-node bifurcation, additional structures in form of downstream streaks together with associated pairs of counter-rotating downstream vortices is added to the solution. The solution thereby grows while its interior structure remains essentially unchanged. 

Both travelling wave snaking branches $TW1$ and $TW2$ are invariant under shift-reflect symmetry. This symmetry is preserved and not broken by the saddle-node bifurcations along the branch. Consequently, the saddle-node bifurcations simultaneously add structures symmetrically at both fronts of the travelling wave. Figure \ref{fig:TW_fields} shows the evolution of the travelling waves in the snaking region for one oscillation. In this range, $TW1$ undergoes a wide oscillation and $TW2$ a narrow oscillation. Although the symmetry relation between $TW1$ and $TW2$ is broken for non-zero suction, for small values of the suction velocity the velocity fields visually still appear as if they were related by rotational symmetry.

Figure \ref{fig:RS_CS_fields} visualises the flow fields of the returning state $RS$ and the connecting state $CS$ branches at the points indicated in figure \ref{fig:one_course}($a$). Each point along both branches corresponds to two velocity fields related by $z$-reflection symmetry $\sigma_z$. We only visualise one of the two symmetry-related velocity fields. Neither the $RS$ nor the $CS$ solutions are invariant under any discrete symmetry. Consequently, the growing and shrinking of the solution along the branch is not symmetric. To visualise changes in the velocity field along the branch, we overlay two successive solutions shown in figure \ref{fig:RS_CS_fields}. Figure \ref{fig:moving_frame} shows the zero contour line of the average streamwise velocity of two consecutive flow fields. The differences are small and located at the fronts. To further highlight the small differences and reveal growing and shrinking of the solution, contours of the amplitude difference of the average streamwise velocity $|\langle u_2 \rangle_x| - |\langle u_1 \rangle_x|$ are visualised. 

The returning state branch $RS$ (thin blue line in figure \ref{fig:one_course}) bifurcates in a pitchfork bifurcation from $TW2$ at $(6a)$. Towards $(6c)$, while decreasing in Re, the solution increases in strength at its left front and weakens on its right. One may describe the localised solutions as a superposition of a periodic pattern with an envelope that supports the internal core structure and damps the exterior towards laminar flow. The change of the solution along $(6a) \to (6c)$ corresponds to a left-shift of the envelope (see figure \ref{fig:moving_frame} (top left panel)). From $(6c) \to (6e)$, close to the next saddle-node bifurcation, the flow field gets stronger at both sides (see figure \ref{fig:moving_frame} (middle left panel)). This segment of the $RS$ branch is the remainder of an equilibrium branch at zero suction. The simultaneous and almost symmetric growth of the solution at both fronts along this segment resembles the growth of the equilibrium solution along the equilibrium branch at zero suction. Finally, from $(6e) \to (6g)$, close to the pitchfork bifurcation off $TW2$, the solution increases in strength on the right and weakens on the left front (figure \ref{fig:moving_frame} (bottom left panel)). The evolution along this part of the $RS$ branch corresponds to a right-shift of the envelope. Along a full $RS$ branch, the envelope of the localised solutions first shifts to the left, then grows symmetrically, and finally shifts to the right. After this sequence, both fronts have grown equally so that the $RS$ branch acquires the symmetry of $TW2$ it bifurcated from. The symmetry related branch $RS' = \sigma_z RS$ bifurcates and reconnects to $TW2$ together with $RS$ but the growth along the branch is inverted: A right-shift is followed by symmetric growth and a final left-shift.

The connecting state $CS$ branch bifurcates from $TW2$ in a pitchfork bifurcation at point $(6b)$. From $(6b) \to (6d)$ the flow fields on the $CS$ branch strengthens on the left and weakens on the right front, which corresponds to a left-shift of the envelope (figure \ref{fig:moving_frame} (top right panel)). From $(6d) \to (6f)$ the flow fields grow almost symmetrically at both fronts (figure \ref{fig:moving_frame} (middle right panel)). This part of the $CS$ branch is the remainder of an equilibrium branch at zero suction. Finally, from $(6f) \to (6h)$ the flow fields again strengthen on the left front and weaken at the right front, corresponding to a left-shift (figure \ref{fig:moving_frame} (bottom right panel)). At point $(6h)$ the $CS$ branch connects to the $TW1$ branch in a pitchfork bifurcation. Along the $CS$ branch the envelope first moves to the left, then grows symmetrically, and finally moves to the left again. 
As a result of this growth sequence involving a net shift to the left, the centrally located low speed streak (a blue streak in figure \ref{fig:RS_CS_fields}) is replaced by the neighbouring high speed streak at the left (a red streak in figure \ref{fig:RS_CS_fields}), which now sits at the center of the localised solution. A centrally located high speed streak is characteristic of $TW1$. Consequently, the $CS$ branch starts on $TW2$ and terminates on $TW1$. The symmetry-related branch $CS' = \sigma_z CS$ bifurcates from $TW2$ and, as $CS$ connects to $TW1$. However, along $CS'$ the solution exhibits a net shift to the right until the solution is no longer centered at a low speed streak but on the next high-speed streak at the right. Since continuous shifts in spanwise direction are part of the system's equivariance group, despite net shifts in opposite directions, both connecting state branches $CS$ and $CS'$ connect to the same $TW1$ branch. 

\section{Discussion}\label{discussion}

Non-vanishing suction velocity $V_s$ causes splitting of the $TW1$ and $TW2$ branches and creates new $CS$ and $RS$ branches from the remainders of $EQ$ and rung branches. In this section we relate these modifications of the snakes-and-ladders bifurcation structure due to the suction to the breaking of symmetries of PCF with zero suction. Moreover, we discuss additional modifications of the bifurcation structure observed at large amplitudes of the suction velocity $V_s$.

\subsection{Snaking solutions and symmetry subspaces in the presence of suction}\label{state_space_symm_break}

\begin{figure}
  \centerline{\includegraphics[width=0.7\textwidth]{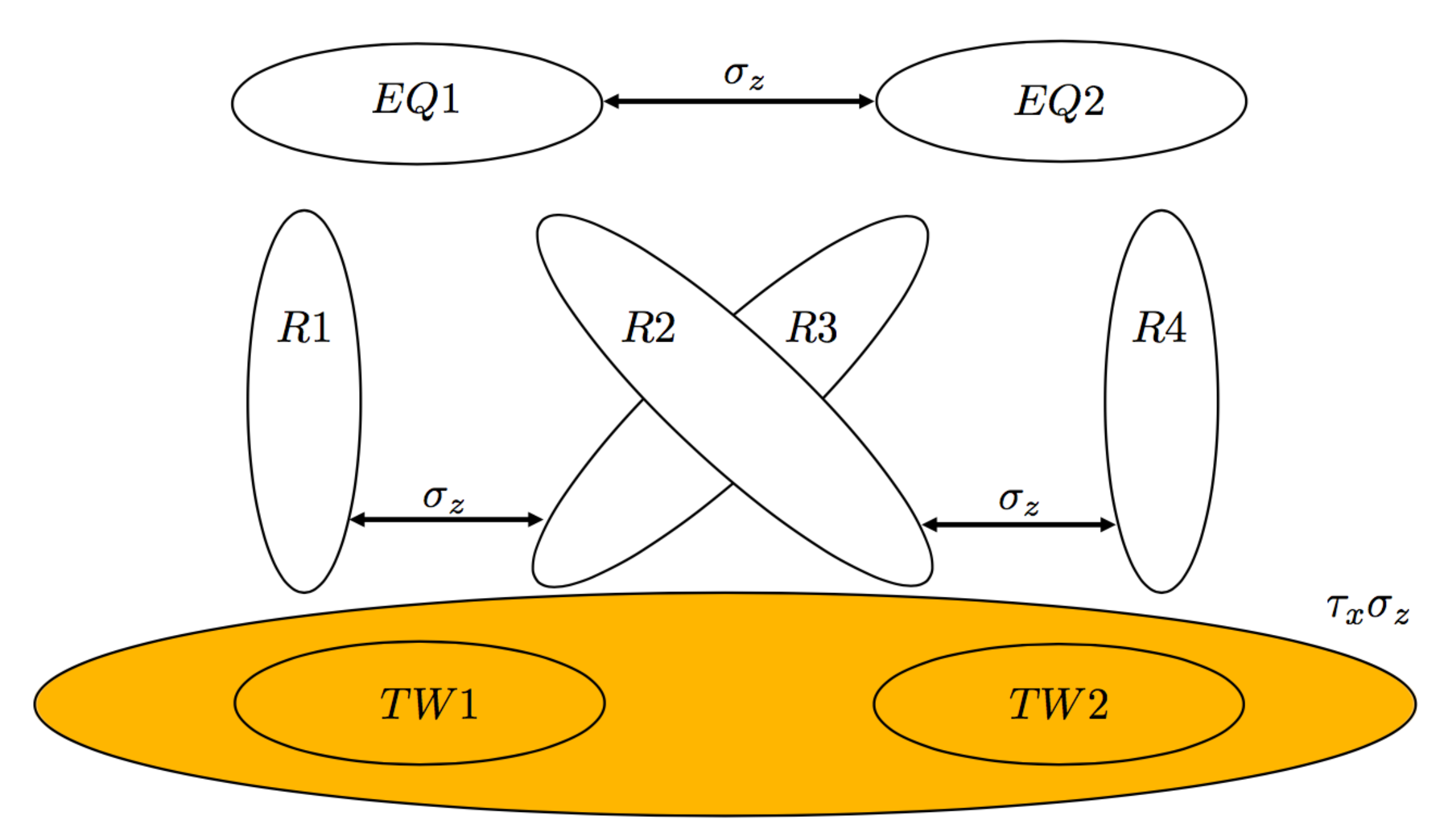}}
  \caption{ 
  Symmetry relations (arrows) between solution branches of plane Couette flow in the presence of non-vanishing wall-normal suction velocity. For $V_s\neq0$, the inversion symmetry relation $\sigma_{xyz}$ between $R1$ and $R2$ and between $R3$ and $R4$, the rotational symmetry relation $\sigma_{xy}$ between $TW1$ and $TW2$ and the inversion symmetry subspace, all present for $V_s=0$ (see figure \ref{fig:schematic_no_suction}) are broken.}
\label{fig:schematic_with_suction}
\end{figure}

Wall normal suction breaks the rotational symmetry of plane Couette flow. As a result, any symmetry subspace or symmetry relation originating from rotational symmetry is not present for non-zero suction velocity. Consequently, for non-zero suction, there is no inversion symmetry subspace, $\sigma_{xyz}$ and states that are inversion-symmetric equilibria at zero suction travel in the streamwise and the spanwise directions. 
Moreover, the rotational symmetry relation between the two travelling wave branches and the inversion symmetry relation between rung states vanishes. Figure \ref{fig:schematic_with_suction} schematically indicates the configuration of symmetry subspaces and symmetry relations of states in the presence of wall-normal suction. Since the symmetry relation between the two travelling wave branches is broken by the wall-normal suction, we expect a separation of the travelling wave branches $TW1$ and $TW2$ in the bifurcation diagram showing dissipation versus Reynolds number. 
Both solution branches that are equilibria at zero suction remain symmetry related, $EQ1 = \sigma_z EQ2$ which implies equal dissipation so that they remain represented by a single curve. At zero suction all four rung states are related by two different symmetry transformations. One of those two symmetries is broken by suction so that the four rungs split into two groups of two symmetry-related branches each, $R1=\sigma_z R3$ and $R2=\sigma_z R4$. Thus two separate curves in the bifurcation diagram represent two symmetry-related solution branches each. Rungs and equilibrium branches at zero suction transform into connecting states $CS$ and returning states $RS$ at non-zero suction. 

\subsection{Front growth controls bulk velocity and oscillation width}\label{bulkVel}

\begin{figure}
  \centerline{\includegraphics[width=0.6\textwidth]{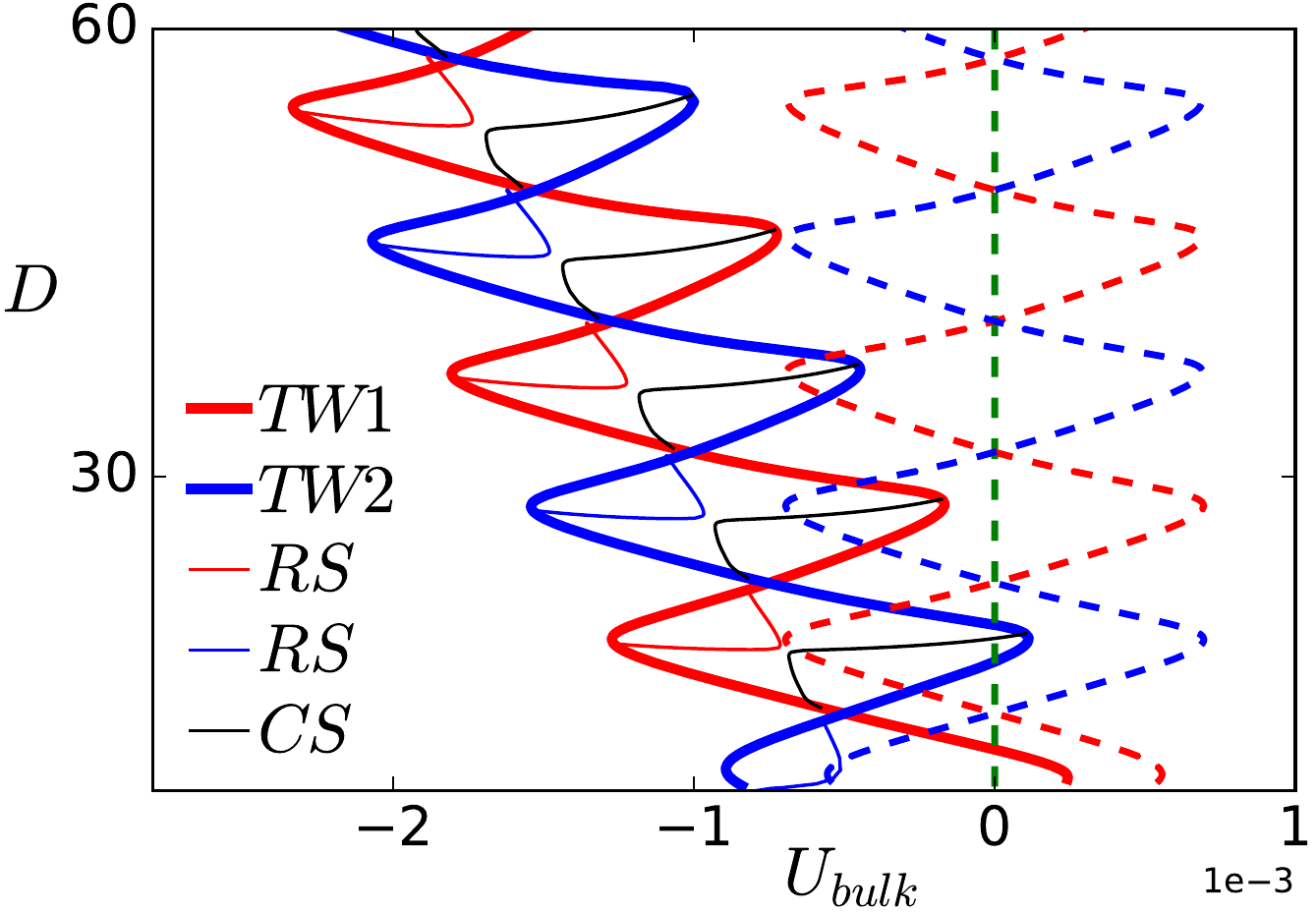}}
  \caption{Dissipation versus bulk velocity in snaking region of plane Couette flow at zero suction (dashed) and for suction velocity $V_s=10^{-4}$. Solution branches are indicated in the legend. At zero suction the bulk velocity of both travelling wave branches oscillates around zero. With finite suction, an additional trend, linear in dissipation and thus size of the solution, is observed.}
\label{fig:bulk_velocity}
\end{figure}

The mean pressure gradient in both spanwise and downstream direction is imposed to be zero and each solution selects its bulk velocity $U_{bulk}$, the $y-z$ averaged streamwise velocity. Figure \ref{fig:bulk_velocity} shows the variation of the bulk velocity of the snaking solutions both for PCF with zero suction and for a suction velocity of $V_s=10^{-4}$. At zero suction, the $EQ$ branch is invariant under $\sigma_{xyz}$, implying zero bulk velocity. The bulk velocity of the travelling wave solutions periodically varies around zero as the solution undergoes snaking. The magnitude of the $U_{bulk}$ oscillations is independent of the spatial extent of the solution which suggests the non-zero bulk velocity is generated by the fronts, at which high- and low-speed streaks are growing symmetrically \cite{Gibson2016}. While low-speed streaks are created, the bulk velocity becomes negative, and when high-speed streaks grow at the fronts the bulk velocity becomes positive.
When suction is applied, the oscillations in bulk velocity are overlaid by an additional negative component that is linearly proportional in the dissipation. The linear dependence on dissipation and thereby size of the solution suggests that the linearly growing component of the bulk velocity is generated by an unchanging internal structure of the solution while the oscillations are due to growth at the fronts. 

The front-mediated oscillations in bulk velocity relative to the linear trend are correlated with the width of snaking oscillations of the travelling wave branches. Excursions to higher bulk velocity are observed when the branch undergoes a wide oscillation, i.e. a wider range of Reynolds numbers (see figure \ref{fig:one_course}). Likewise, narrow oscillations are observed when the bulk velocity approaches local minima. Maxima of the bulk velocity are linked to high-speed streaks growing at the front while for minima in the bulk velocity, growing low-speed streaks are observed. Since $TW1$ remains related to $TW2$ by the approximate though broken rotational symmetry $\sigma_{xy}$, a growth of high-speed streaks along the $TW1$ branch implies the growth of low-speed streaks along $TW2$ and vice versa. The growth of high- and low-speed streaks of approximately symmetry-related travelling wave branches thus explains why wide and narrow oscillations of $TW1$ and $TW2$ both alternate and moreover occur such that a wide oscillation of $TW1$ coincides with a narrow one of $TW2$ and vice versa.

\subsection{Splitting of the travelling waves}\label{linear}

\begin{figure}
  \center{
  ($a$)
  \includegraphics[width=0.47\textwidth]{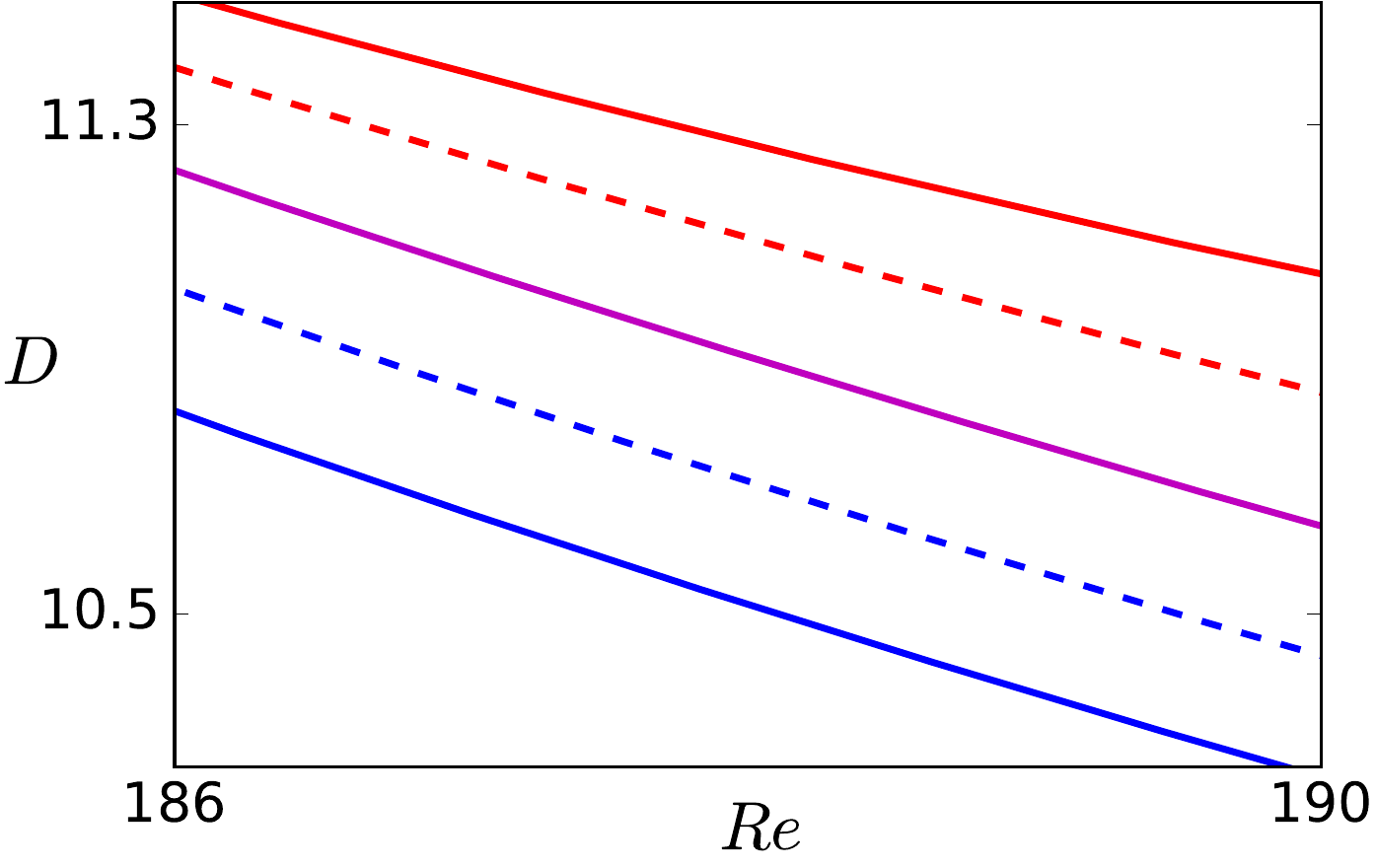}
  ($b$)
  \includegraphics[width=0.44\textwidth]{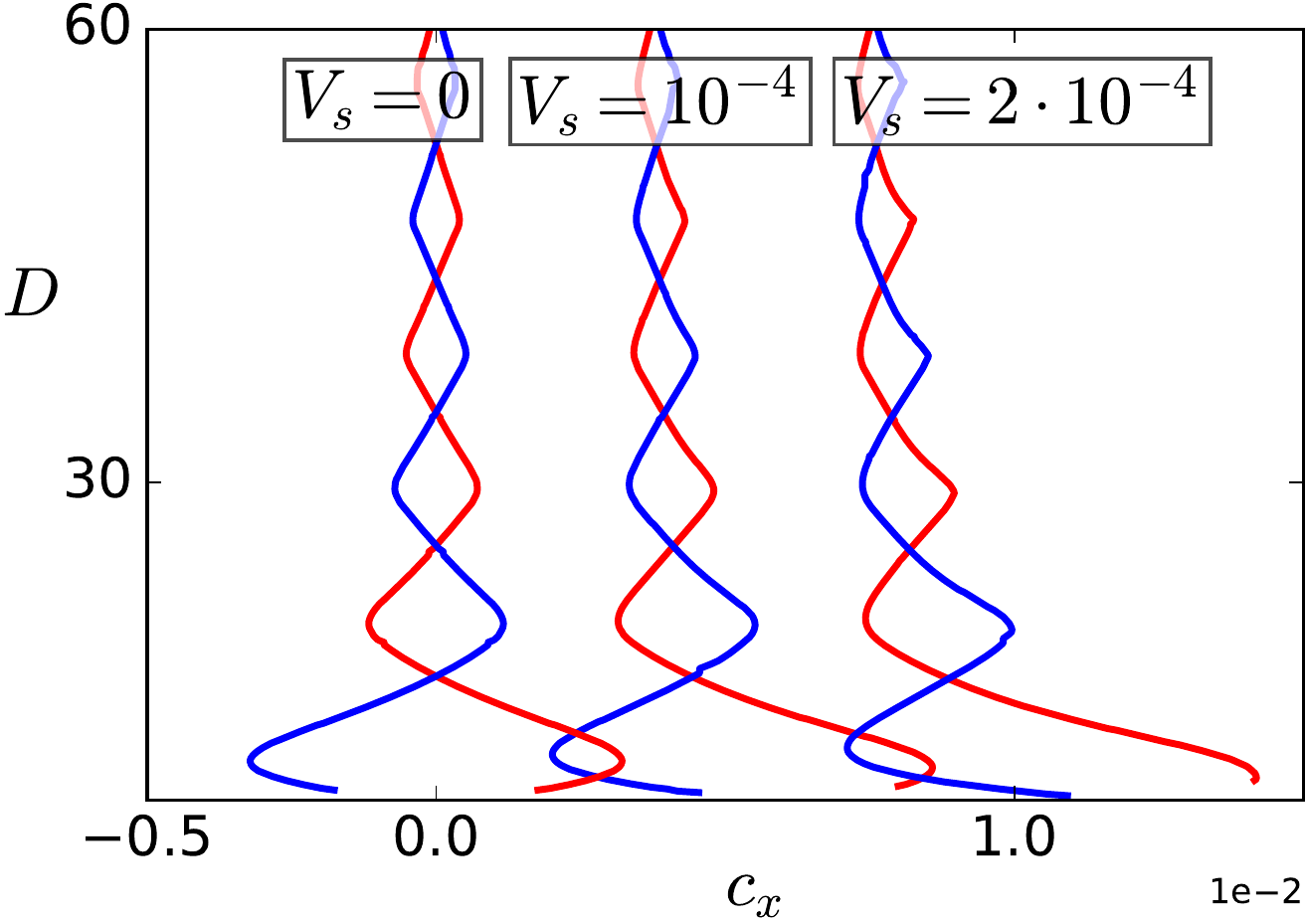}
  }
  \caption{
  $(a)$ Bifurcation diagram of $TW1$ (red and magenta) and $TW2$ (blue and magenta) representing dissipation $D$ as a function of Reynolds number $Re$ for different values of the wall-normal suction velocity, $V_s=0$ (magenta line), $V_s=10^{-4}$ (dashed lines) and $V_s=2\cdot10^{-4}$ (solid lines). 
  For non-zero suction velocity, the branches of $TW1$ and $TW2$, which are represented by a single curve for $V_s=0$ (magenta), split symmetrically. They are located at the same distance but on opposite sides of the curve with $V_s=0$.
  $(b)$ Dissipation $D$ versus streamwise wave speed $c_x$ of $TW1$ (red) and $TW2$ (blue) for the same values of the wall-normal suction velocity as in part $(a)$. The values of the suction velocity are indicated on the curves in the figure. The wave speed of both $TW1$ and $TW2$ is shifted by equal amounts. The shift is proportional to the suction velocity $V_s$.
  }
\label{fig:splitting}
\end{figure}

In PCF without suction, both travelling wave solution branches, $TW1$ and $TW2$, are represented by a single curve in the bifurcation diagram in terms of $D$ versus $Re$. Wall-normal suction breaks the rotational symmetry of the system $\sigma_{xy}$ and results in splitting of the travelling wave solution branches (see figure \ref{fig:snaking_with_suction}). A subset of the bifurcation diagram is enlarged in figure \ref{fig:splitting}($a$), where dissipation $D$ as a function of $Re$ is presented for $TW1$ and $TW2$. The branches are shown for three values of the wall-normal suction velocity, $V_s=0$, $V_s=10^{-4}$ and $V_s=2\cdot10^{-4}$. 
For non-zero suction, branches split so that $TW1$ and $TW2$ curves are symmetrically located around the zero-suction curve, with equal distance but on opposing sides. The distance between both split curves grows linearly with the suction velocity. To rationalise this splitting behaviour we consider modifications of the solution of the governing equations at leading, namely linear, order in $V_s$. At zero wall-normal suction $V_s=0$, the laminar flow solution is a linear profile $\mathbf{U}(y)=y\hat{e}_x$. Non-zero wall suction modifies the travelling wave solutions $TW1$ and $TW2$. When decomposing the velocity into a laminar flow and deviations, this modification affects both the laminar base flow $\mathbf{U}$ and the deviation from the modified laminar base flow $\mathbf{u}$, as we will describe in this section.

\subsubsection{Relation between travelling waves for inverted suction velocity}
Solutions cannot only be followed from zero to positive suction velocity $V_s=+|V_s|$ but also to negative values $V_s=-|V_s|$. The latter physically implies blowing through the wall. Since the rotational symmetry $\sigma_{xy}$, relating $TW1$ to $TW2$ at $V_s=0$ and broken by suction, transforms a flow with suction to a flow with equal amplitude blowing, the travelling wave branches for inverted suction velocity are related by 
\begin{align}\label{positive_negative_Vs}
TW2 \big|_{V_s}=\sigma_{xy}\; TW1 \big|_{-V_s}\;.
\end{align}

\subsubsection{The modification of the laminar solution}
At linear order in $V_s$ the laminar solution reads
\begin{align*}
\mathbf{U} = \left(y+\frac{Re V_s}{2}\left(1-y^2\right)\right)\hat{e}_x -V_s\hat{e}_y.
\end{align*}
Consequently, the modification of the laminar solution with respect to the laminar solution of PCF for zero suction velocity $\mathbf{U}'=\mathbf{U}-\mathbf{U}_{V_s=0}$ is
\begin{align}\label{base_mod}
\mathbf{U}' =  V_s\left[\frac{\Rey}{2}\left(1-y^2\right)\hat{e}_x -\hat{e}_y\right] = V_s\mathbf{\bar{U}}(y,\Rey),
\end{align}
where $\mathbf{\bar{U}}$ is a function of Reynolds number and the wall-normal coordinate $y$ only. $\mathbf{\bar{U}}$ is independent of the suction velocity so that the laminar flow modification $\mathbf{U}'$ is linearly proportional to the suction velocity $V_s$.

\subsubsection{The modification of the deviation from the laminar solution}
The travelling wave solutions $TW1$ and $TW2$ are equilibria in a moving frame of reference that translates at their specific wave speeds $c_x$ in the $\hat{e}_x$ direction with respect to the lab frame. The travelling wave solutions thus satisfy the condition
\begin{align*}
(\mathbf{u}_t-c_x \,\hat{e}_x)\cdot\nabla \mathbf{u}_t=-\nabla p+\frac{1}{Re} \nabla^2 \mathbf{u}_t
\end{align*}
where $c_x$ is the wave speed associated with the total flow field $\mathbf{u}_t = \mathbf{U}(y)+\mathbf{u}(x,y,z)$. 
The solution of this equation $(\mathbf{u}_t, p,c_x)$ may be decomposed into three parts: the solution for zero suction, a modification of the laminar solution due to suction, and the modification of the deviation from the laminar solution:
\begin{align}\label{decomposition}
\begin{cases}
\mathbf{u}_t & = \mathbf{u}_0 +  \mathbf{U}' + \mathbf{u}'\\
p & = p_0 + p'\\
c_x & = c_{x0} + c'_x,
\end{cases}
\end{align}
where $(\mathbf{u}_0, p_0,c_{x0})$ denotes the total solution for zero suction, $\mathbf{U}'$ is the modification of the laminar solution, and $(\mathbf{u}', p',c'_x)$ is the modification of the deviation from the laminar solution. For small suction velocity, quadratic interactions among $\mathbf{U}'$, $\mathbf{u}'$ and $c'_x \hat{e}_x$ resulting from the inertial term are neglected. At leading order, the condition for a travelling wave solution at small suction thus reads
\begin{align*}
\mathbf{u}_0\cdot\nabla \mathbf{U}'&+\mathbf{u}_0\cdot\nabla \mathbf{u}'+\mathbf{U}'\cdot\nabla \mathbf{u}_0+\mathbf{u}'\cdot\nabla \mathbf{u}_0-c_{x0}\cdot \nabla \mathbf{U}'-c_{x0}\cdot \nabla \mathbf{u}'-c'_x\cdot \nabla \mathbf{u}_0\\
&=-\nabla p'+\frac{1}{Re}\nabla^2 \mathbf{U}'+\frac{1}{Re}\nabla^2 \mathbf{u}'.
\end{align*}
Rearranging using the base flow modifications at linear order in $V_s$ (Equation \ref{base_mod}), yields
\begin{align}\label{eq-split-01}
\nonumber V_s\bigg[\mathbf{u}_0\cdot\nabla \mathbf{\bar{U}}&+\mathbf{\bar{U}}\cdot\nabla \mathbf{u}_0-c_{x0}\cdot \nabla \mathbf{\bar{U}}-\frac{1}{Re}\nabla^2 \mathbf{\bar{U}}\bigg]\\
&=-\nabla p'+\frac{1}{Re}\nabla^2 \mathbf{u}'-\mathbf{u}_0\cdot\nabla \mathbf{u}'-\mathbf{u}'\cdot\nabla \mathbf{u}_0+c_{x0}\cdot \nabla \mathbf{u}'+c'_x\cdot \nabla \mathbf{u}_0,
\end{align}
where the suction velocity has been isolated. Only the primed variables depend on $V_s$ so that the entire left-hand side is proportional to $V_s$. The equation determines $(\mathbf{u}', p',c'_x)$ as a function of $V_s$. Since the operators acting on the primed variables are linear (right-hand side), the solution $(\mathbf{u}', p',c'_x)$ is linearly proportional to the suction velocity $V_s$. 

Dissipation of $TW1$ with wall-normal suction velocity $V_s$ can be expressed as
\begin{align*}
    D=D\big|_{V_s=0}+\frac{1}{2L_x}\int_{-L_z/2}^{L_z/2}\int_{-L_x/2}^{L_x/2} \left(\frac{\partial \mathbf{u}'}{\partial y}\bigg|_{y=-1}+\frac{\partial \mathbf{u}'}{\partial y}\bigg|_{y=1}\right) dx dz
\end{align*}
where $D\big|_{V_s=0}$ is the dissipation of both $TW1$ and $TW2$ at zero suction velocity. According to relation \ref{positive_negative_Vs}, the dissipation of $TW2$ with suction velocity $V_s$ is equal to the dissipation of $TW1$ for inverted suction velocity ($-V_s$). As a result, the dissipation of $TW2$ follows from the same expression when the suction velocity in Equation \ref{eq-split-01} is inverted ($V_s \to -V_s$). Equation \ref{eq-split-01} implies $\mathbf{u}' (-V_s) = -\mathbf{u}'(V_s)$. Consequently, the dissipation of $TW2$ with wall-normal suction velocity $V_s$ is given by
\begin{align*}
    D=D\big|_{V_s=0}-\frac{1}{2L_x}\int_{-L_z/2}^{L_z/2}\int_{-L_x/2}^{L_x/2} \left(\frac{\partial \mathbf{u}'}{\partial y}\bigg|_{y=-1}+\frac{\partial \mathbf{u}'}{\partial y}\bigg|_{y=1}\right) dx dz\;.
\end{align*}
Suction thus changes the dissipation of travelling wave solutions $TW1$ and $TW2$ relative to the zero suction case by amounts that are equal in magnitude but opposite in sign.

Equation \ref{eq-split-01} implies that the entire solution $(\mathbf{u}', p',c'_x)$ is linear in $V_s$ so that also modifications of the wave speed $c'_x$ are proportional to the suction velocity. 
Equation \ref{positive_negative_Vs} implies that the modification of the wave speed of $TW2$ with wall-normal suction velocity $V_s$ is equal to \emph{minus} the modification of the wave speed of $TW1$ for inverted suction velocity ($-V_s$). Together with the linear dependence of wave speed modifications $c'_x$ on $V_s$, this implies that the modifications of the wave speed are equal for both $TW1$ and $TW2$. Figure \ref{fig:splitting}($b$) presents the bifurcation diagram of $TW1$ and $TW2$, in terms of dissipation $D$ versus wave speed $c_x$, for different values of the wall-normal suction velocity, $V_s=0$, $V_s=10^{-4}$ and $V_s=2\cdot10^{-4}$. This data confirms that small values of the wall-normal suction velocity $V_s$ change the wave speeds of both $TW1$ and $TW2$ by the same amount. The amount is linearly proportional to the suction velocity.

\subsection{Alternating bifurcations of $CS$ and $RS$ off travelling wave branches}\label{periodicity}

\begin{figure}
  \centerline{\includegraphics[width=0.5\textwidth]{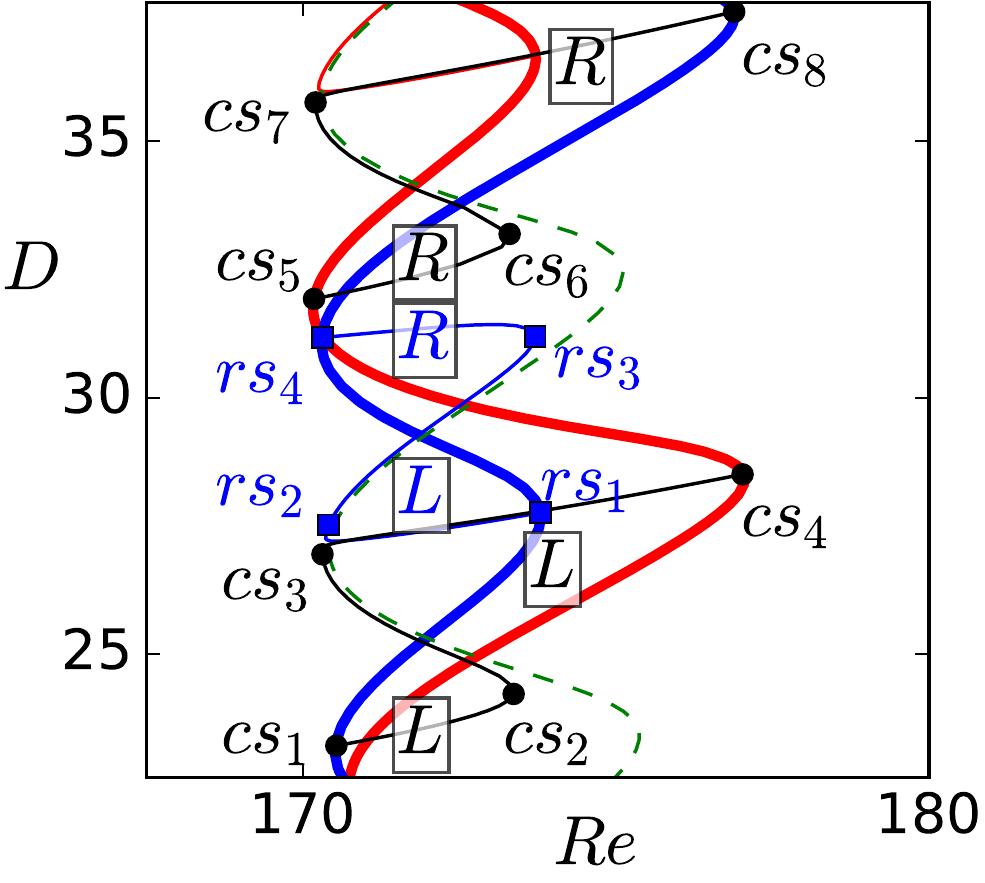}}
  \caption{
  Bifurcation diagram representing $D$ versus $Re$ showing both the $TW1$ (thick red) and $TW2$ (thick blue) branches, together with two successive $CS$ branches (thin black) and an $RS$ branch (thin blue) for $V_s=10^{-4}$. For reference, also the curve representing $EQ1$ and $EQ2$ for $V_s=0$ is shown (dashed green line). 
  Along the $CS$ and $RS$ branches that inherit segments from $EQ1$ for $V_s=0$, the direction of the shifts of the envelope of the localised solutions are indicated in the figure. The symmetry-related counterparts of these $CS$ and $RS$ solution branches, that contain remainders from $EQ2=\sigma_{z} EQ1$ for $V_s=0$, shift in opposite directions.}
  
\label{fig:LR_argument}
\end{figure}

Both travelling wave branches, $TW1$ and $TW2$ undergo alternating wide and narrow oscillations in the dissipation versus Reynolds number bifurcation diagram. Along a wide oscillation a pair of high-speed streaks is added at the fronts of the solution, while along a narrow oscillation a pair of low speed streaks is added. After an entire period consisting of a successive wide and narrow oscillation, the fronts connecting the internal periodic pattern of the travelling waves to laminar flow recover their original structure and the solution has grown by two additional pairs of high- and low-speed streaks. The solution thus only grows at the fronts while its internal structure remains unchanged. As a result of this front-driven growth process, the critical Reynolds numbers of corresponding saddle-node bifurcations along successive wide or narrow oscillations line up. The spatial growth of the solution indicated by increasing values of dissipation $D$ continues periodically. 

The growth of the travelling wave branches is driven by saddle-node bifurcations with neutral modes that are localised at the solution fronts \citep{Schneider2010}. Likewise, the pitchfork bifurcations close to the saddle-nodes that create the $RS$ and $CS$ have neutral modes acting only at the fronts. As a result, the entire bifurcation structure, including the bifurcating $RS$ and $CS$ branches, repeats periodically after a pair of oscillations. Moreover, the front structure of $TW1$ reflects that of the $TW2$ branch shifted by half a period. Since all bifurcations are driven at the fronts, the equivalent front structure of both travelling wave branches implies that all branches bifurcating off $TW1$ have corresponding branches bifurcating off $TW2$. Consequently, the entire bifurcation structure of $TW1$ including bifurcations off the branch corresponds to that of the $TW2$ branch shifted by half a period.  

As discussed in section \ref{sec:flowfields_with_Vs}, $CS$ and $RS$ branches are characterised by the sequence of directions in which the envelope of the localised solutions is shifted as one marches along the branch. Starting from the pitchfork bifurcations off a travelling wave at lower $D$ where the branches emerge towards the pitchfork at higher $D$ where the branches terminate on a travelling wave, both $CS$ and $RS$ encounter two envelope shifts. The connecting state $CS$ branch is characterised by two shifts to the same directions either to the right ($+\hat{e}_z$) or to the left ($-\hat{e}_z$). Two shifts in opposing directions (left after right or vice versa) characterise the returning state $RS$ branch.
Figure \ref{fig:LR_argument} enlarges a part of the bifurcation diagram showing both $TW1$ and $TW2$ branches, two successive $CS$ branches and one $RS$ branch connected to $TW2$. Note that each $CS$ or $RS$ curve in the bifurcation diagram represents a pair of $z$-reflection-related branches of $CS$ or $RS$. 
In the figure, we indicate the shifts of the envelope for those $CS$ and $RS$ branches that inherit segments from $EQ1$ for $V_s=0$ (from $cs_{2} \to cs_{3}$, from $rs_{2} \to rs_{3}$ and from $cs_{6} \to cs_{7}$). The symmetry-related counterparts of these $CS$ and $RS$ branches contain remainders from $EQ2$. Due to the symmetry relation, any shift of the branch shown in the figure corresponds to a shift in the opposite direction for the (not shown) symmetry-related branch. If one $CS$ branch twice shifts to the left (from $cs_{1} \to cs_{4}$), the $CS$ for the next higher dissipation (from $cs_{5} \to cs_{8}$) twice shifts in the opposite, here right, direction. 

For $V_s=0$, the two rung state branches that bifurcate off an equilibrium branch in a pitchfork bifurcation are related by $\sigma_{xyz}$. 
For non-zero but small values of $V_s$, the remainders of these two rung state branches form segments of the $CS$ and $RS$ branches (in the figure, $cs_{3} \to cs_{4}\; /\; rs_{1} \to rs_{2}$ and $rs_{3} \to rs_{4}\; /\; cs_{5} \to cs_{6}$). The solutions on these remainders of the rung state branches remain visually almost symmetry-related at small $V_s$. 
As a result of the approximate inversion symmetry, along these two segments of the $CS$ and $RS$ branches, the directions of the shifts of the envelope has to be identical.
Consequently, the shifts along segments of the $RS$ branches are slaved to the shifts of the $CS$ branches along the corresponding segments by the broken inversion symmetry relation of the rung states. The non-symmetric solution branch between two $CS$ branches in the bifurcation diagram must thus be characterised by shifts in two opposing directions and is therefore an $RS$ branch. Symmetry thus implies that $CS$ and $RS$ branches alternate in the bifurcation diagram.

\subsection{Snaking breakdown}\label{breakdown}

\begin{figure}
  \center{
  ($a$)
  \includegraphics[width=0.5\textwidth]{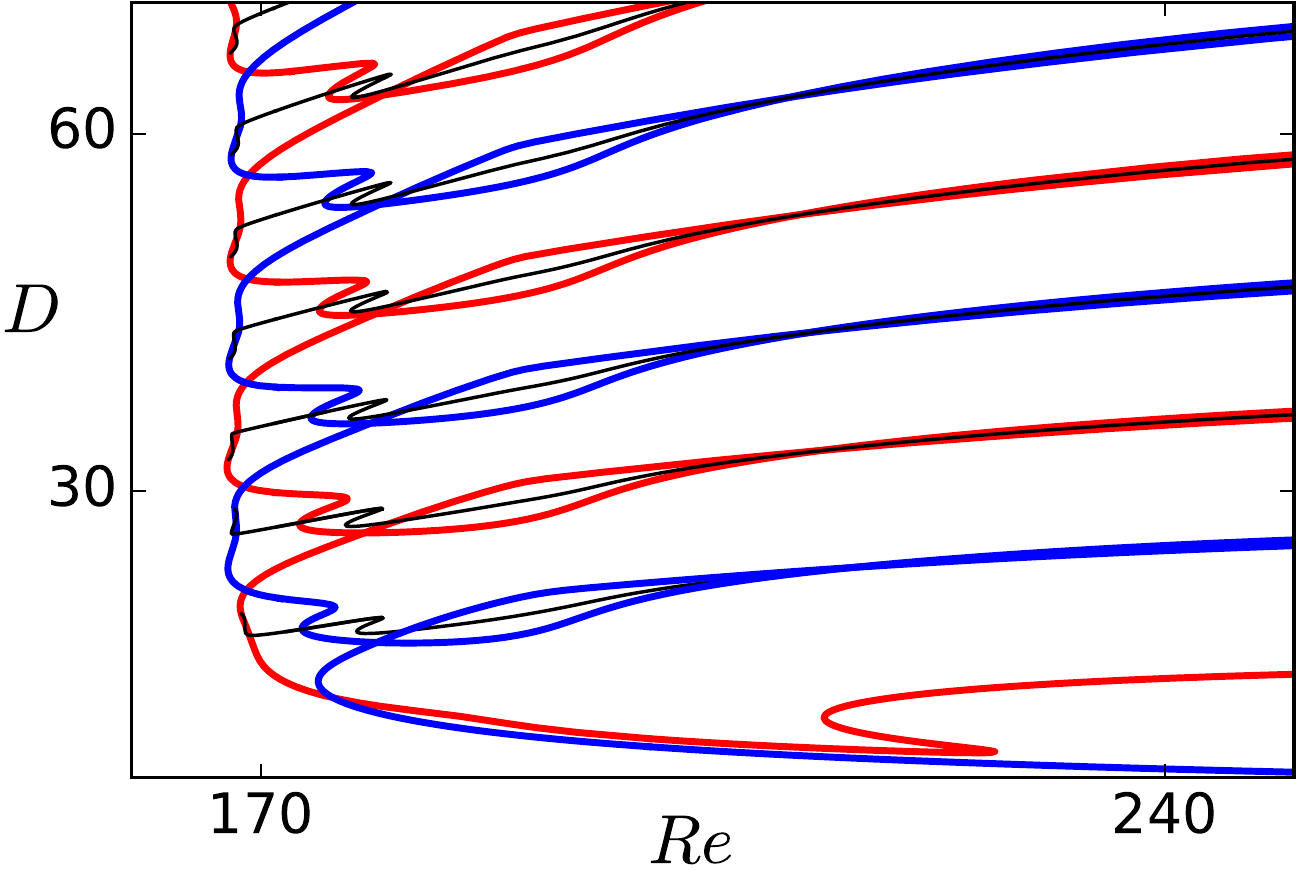}\\
  ($b$)
  \includegraphics[width=0.5\textwidth]{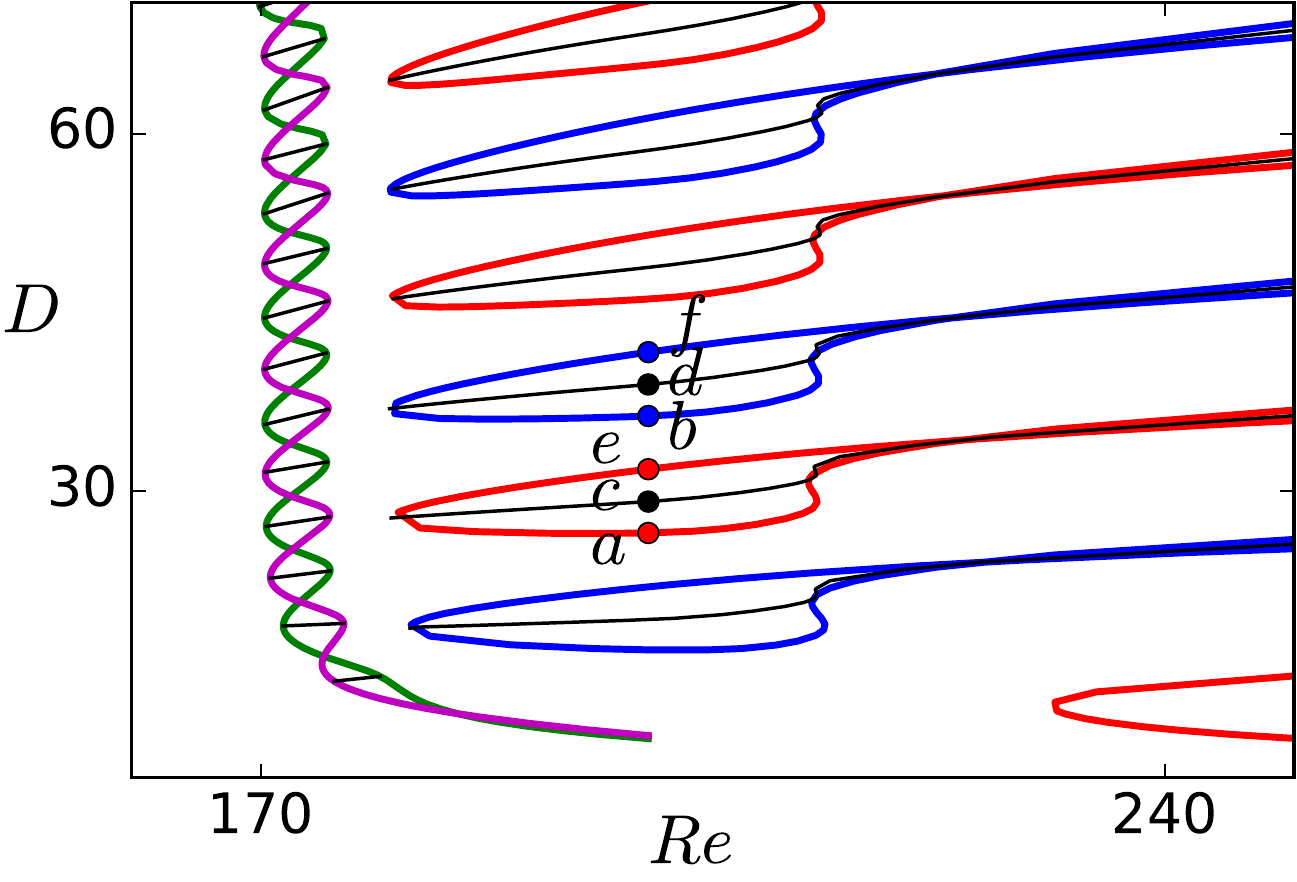}}
  \caption{
  Bifurcation diagram representing $D$ versus $Re$ for ($a$) $TW1$ (red), $TW2$ (blue) and $CS$ (black) at $V_s=6\cdot10^{-4}$; and ($b$) $TW1$- (red), $TW2$- (blue) and $CS$-connected (black) branches at $V_s=0$. At $V_s=0$, the bifurcation diagram of the snaking equilibrium branches (green), travelling waves (magenta) and rung states (black) are shown. The flow fields at the indicated points are visualised in figure \ref{fig:breakdown_fields}. 
  At $V_s=6\cdot10^{-4}$, the snaking branches of $TW1$ and $TW2$ are broken, and $TW1$, $TW2$ and $CS$ branches are merged with solution branches that for $V_s=0$ are separated from the snakes-and-ladders bifurcation structure and extend to high Reynolds numbers.}
\label{fig:breakdown}
\end{figure}

\begin{figure}
    \centering
    ($a$)
    \subfloat{{\includegraphics[width=0.45\textwidth]{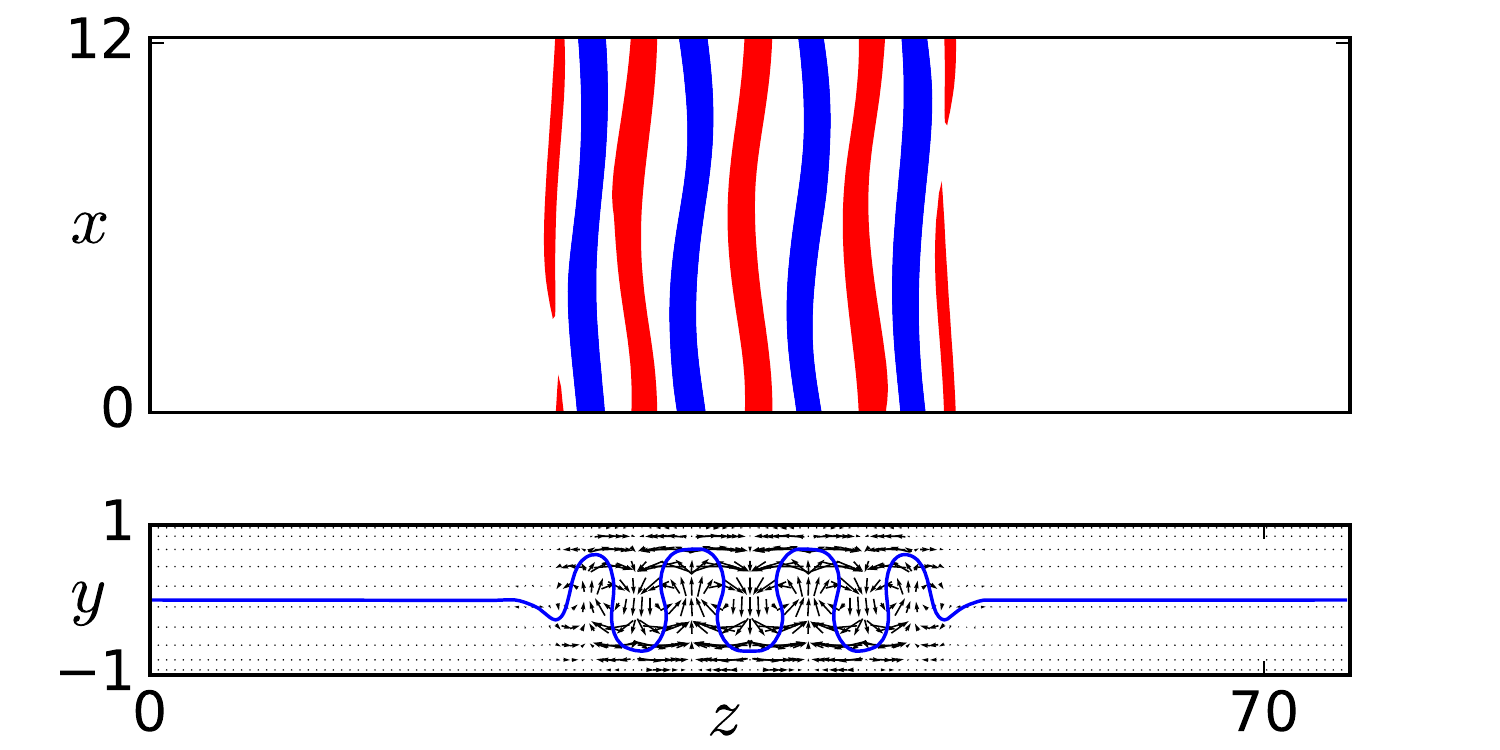} }}
    ($b$)
    \subfloat{{\includegraphics[width=0.45\textwidth]{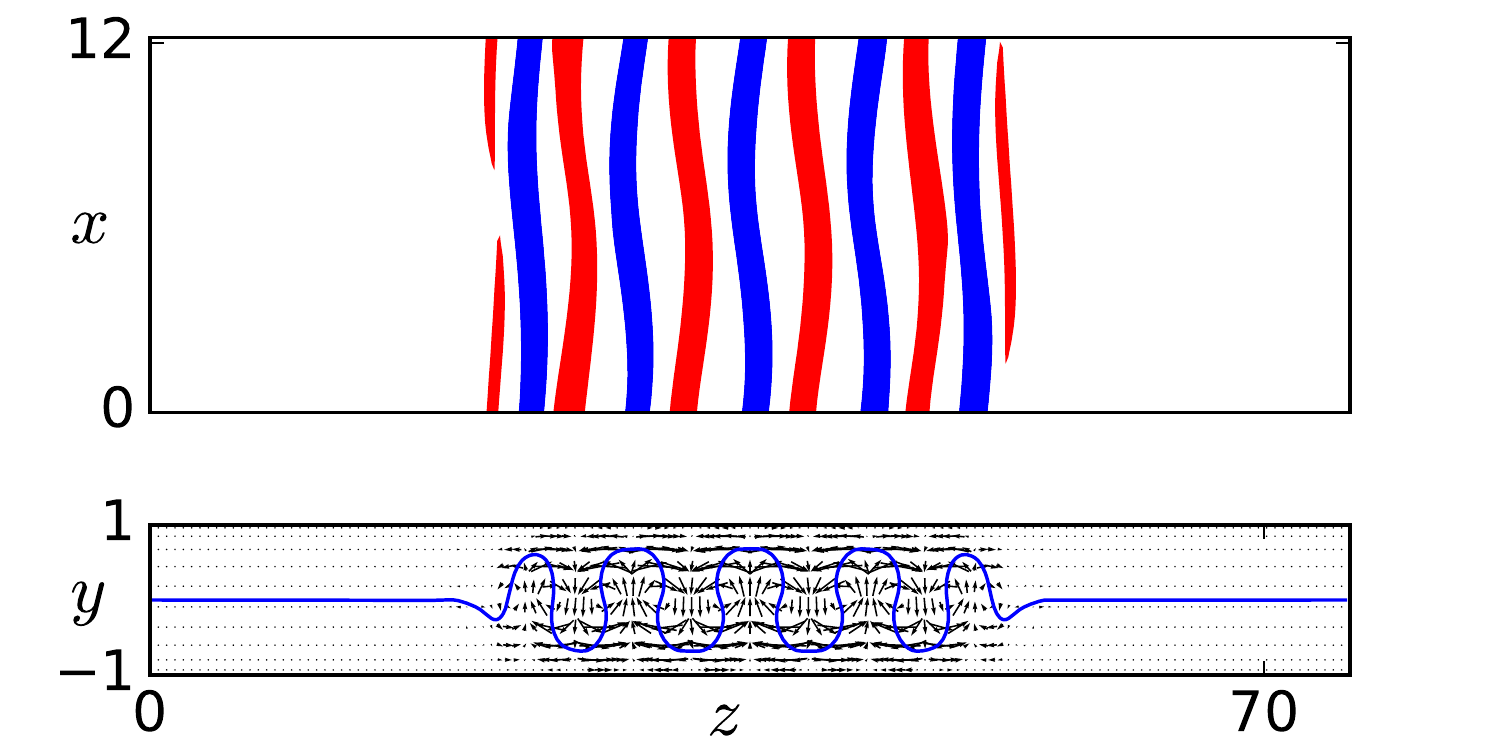} }}\\
    ($c$)
    \subfloat{{\includegraphics[width=0.45\textwidth]{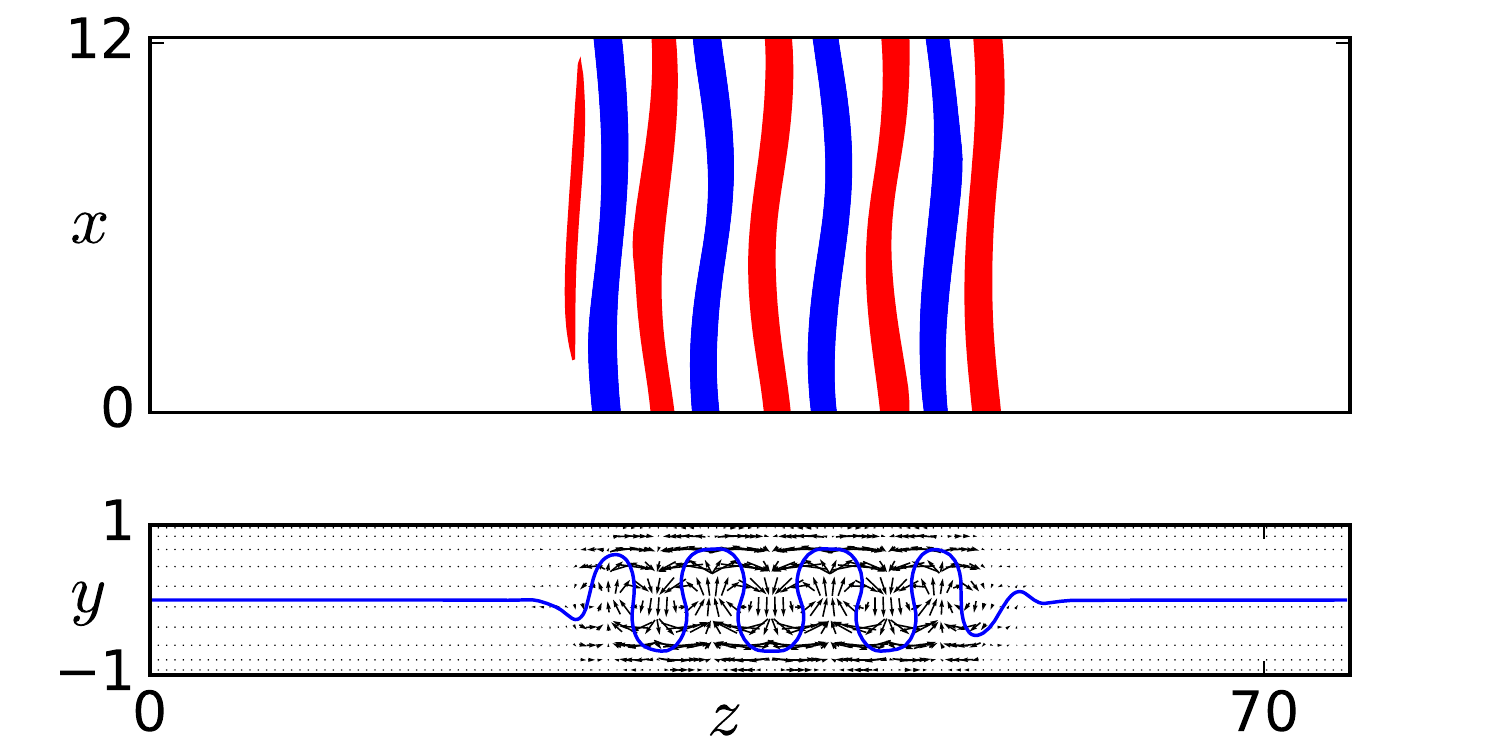} }}
    ($d$)
    \subfloat{{\includegraphics[width=0.45\textwidth]{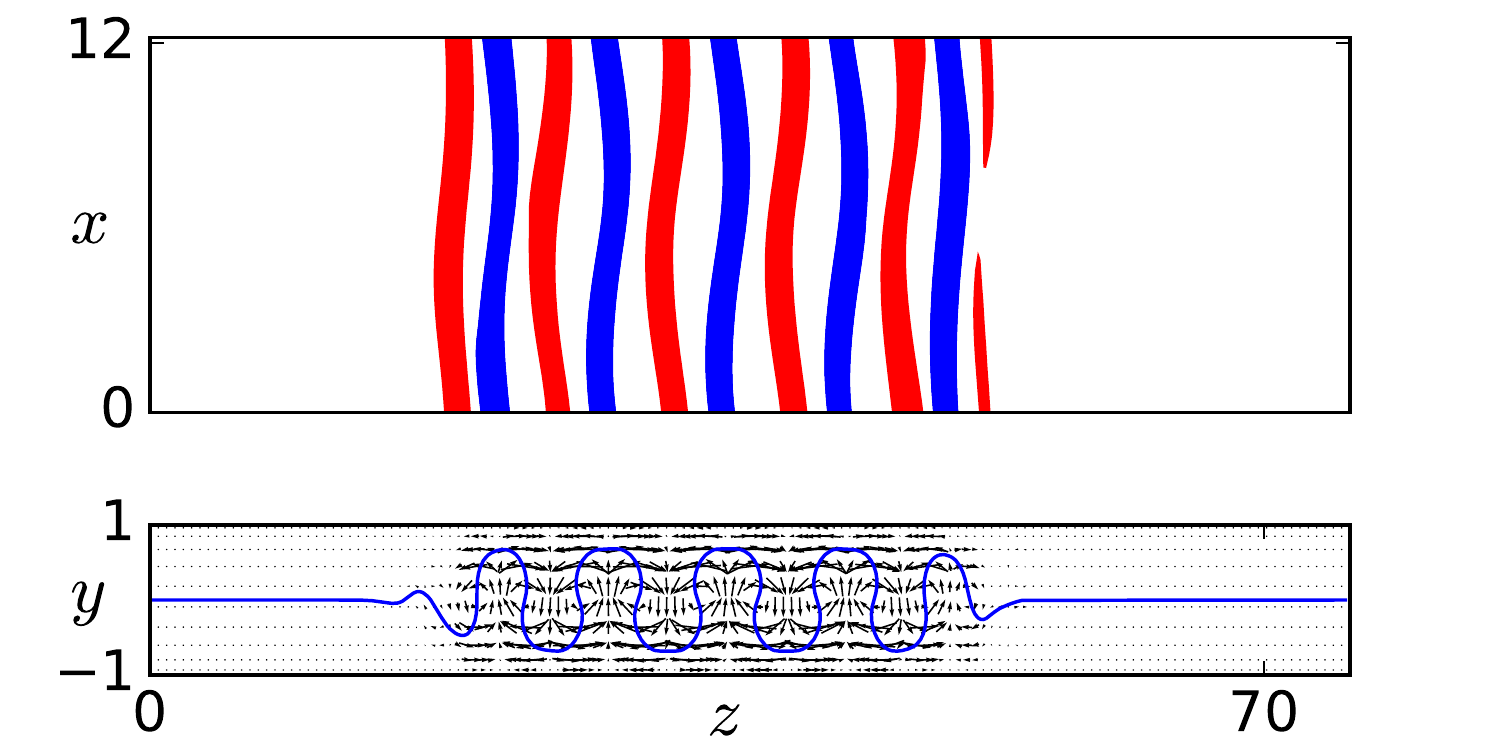} }}\\
    ($e$)
    \subfloat{{\includegraphics[width=0.45\textwidth]{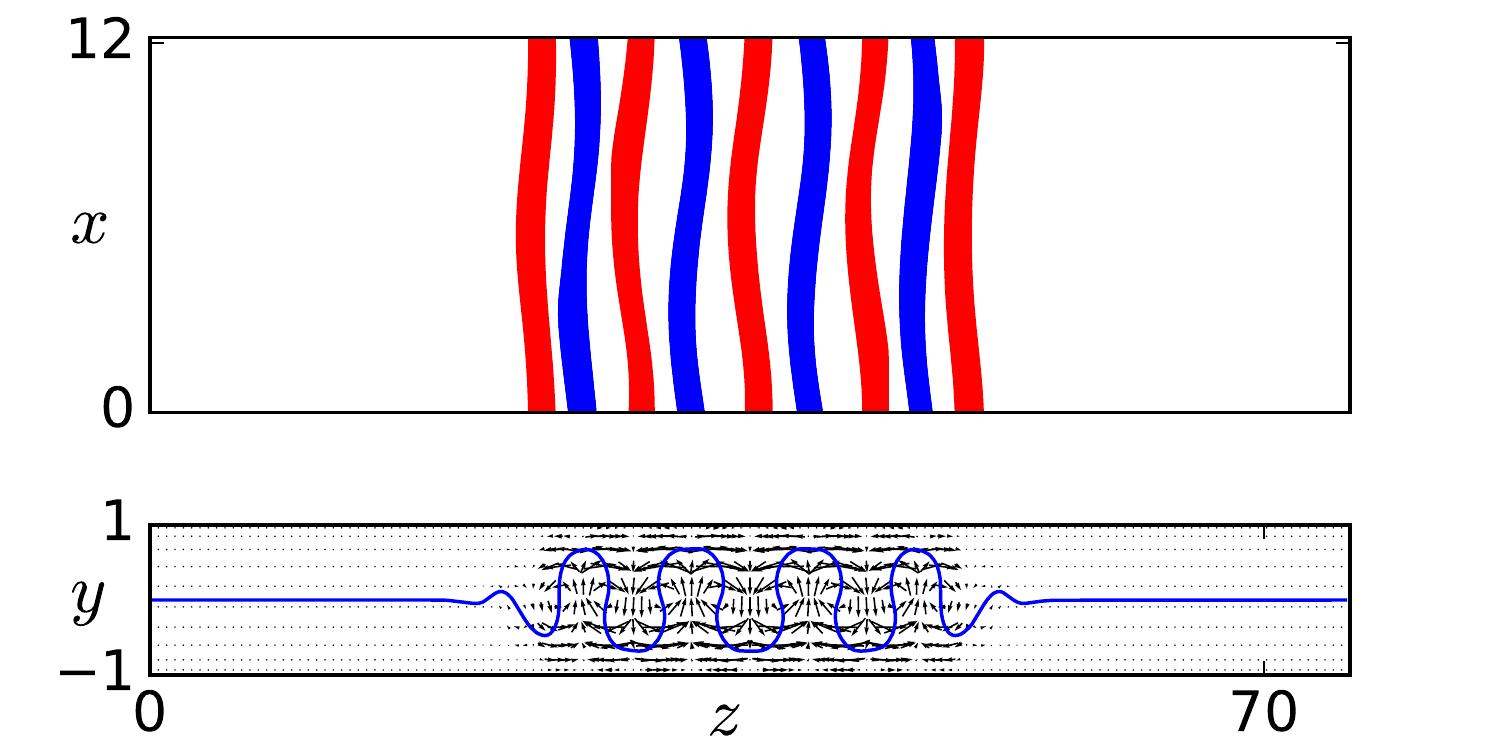} }}
    ($f$)
    \subfloat{{\includegraphics[width=0.45\textwidth]{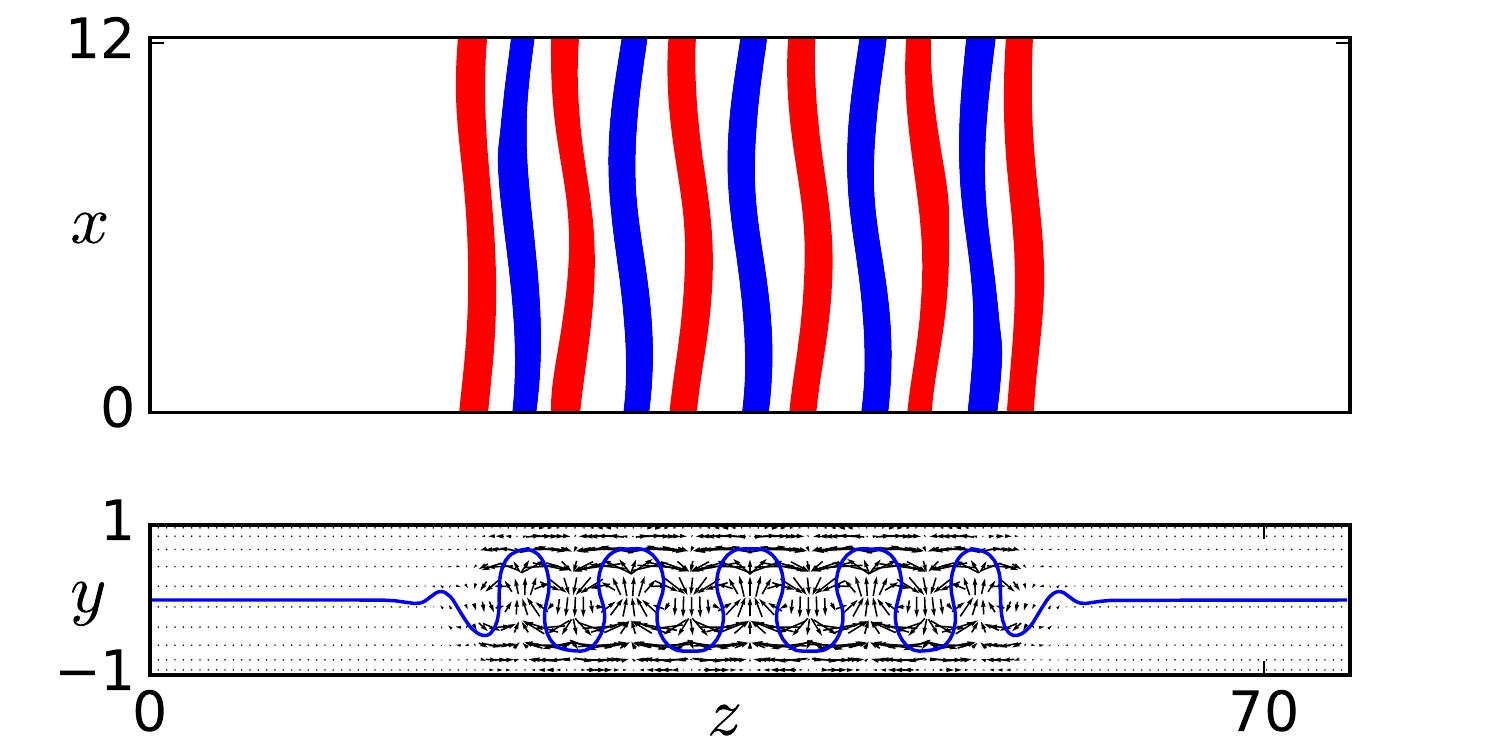} }}
    \caption{Flow fields of plane Couette ($V_s=0$) solutions at points indicated in figure \ref{fig:breakdown}($b$). 
    The flow fields in the left panels are located on the $TW1$-connected branch (panels $a$ and $e$) and on the $CS$-connected branch that bifurcates from the $TW1$-connected branch (panel $c$). The right panels visualise the flow fields that are located on the $TW2$-connected branch (panels $b$ and $f$) and on the $CS$-connected branch that bifurcates from the $TW2$-connected branch (panel $d$).
    The flow fields are visualised as in figure \ref{fig:snaking_with_symms}.}
    \label{fig:breakdown_fields}
\end{figure}

The effect of suction on plane Couette snaking solutions was studied for small suction velocities up to $V_s=2\cdot10^{-4}$. At larger suction velocities, the bifurcation structure fundamentally changes. Figure \ref{fig:breakdown}($a$) shows the bifurcation diagram of $TW1$, $TW2$, and $CS$ branches for $V_s=6\cdot10^{-4}$. The range in $Re$ for which $RS$ branches exist shrinks with increasing $V_s$ (see figure \ref{fig:snaking_with_suction}) and at $V_s=6\cdot10^{-4}$, $RS$ branches are not computed. 
At $V_s=6\cdot10^{-4}$, the snaking travelling wave solution branches, $TW1$ and $TW2$, have broken into multiple disconnected travelling wave branches that reach high Reynolds numbers far beyond the snaking range at $V_s=0$. The two symmetry-related $CS$ branches no longer connect $TW1$ and $TW2$. Instead each $CS$ branch emerging in pitchfork bifurcations off each disconnected travelling wave branch now extends to high Reynolds numbers but no longer appears to terminate in a second pitchfork bifurcation. 

Parametric continuation down in $V_s$ starting from all branches identified at $V_s=6\cdot 10^{-4}$ including those reaching high Reynolds numbers yields previously unknown localised solution branches of PCF without suction. Figure \ref{fig:breakdown}($b$) shows the bifurcation diagram with the characteristic snakes-and-ladders structure together with these newly-found localised solution branches for $V_s=0$. The non-snaking localised solutions include branches that for $V_s=6\cdot 10^{-4}$ are merged with the $TW1$ branch (hereafter referred to as $TW1$-connected), the $TW2$ branch (hereafter referred to as $TW2$-connected), and the $CS$ branches (hereafter referred to as $CS$-connected). The solutions of $TW1$- and $TW2$-connected branches are invariant under the action of the shift-reflect symmetry, $\tau_x\sigma_z$. The solutions of the $CS$-connected branches are not invariant under any discrete symmetry of the system. Instead there are pairs of $CS$-connected solution branches related by $\sigma_z$. Each pair bifurcates from either a $TW1$- or $TW2$-connected branch in a pitchfork bifurcation. Snaking solutions in PCF at $V_s=0$ only exist in a small Reynolds number range of approximately $169 < Re < 177$. The localised non-snaking solution branches, $TW1$-, $TW2$- and $CS$-connected, span a much wider range in Reynolds number (see figure \ref{fig:breakdown}($b$)).

Figure \ref{fig:breakdown_fields} visualises the flow fields of $TW1$-, $TW2$- and two successive $CS$-connected branches at the points indicated in figure \ref{fig:breakdown}($b$).
The internal periodic part and the front structures of the lower branch $TW1$- and $TW2$-connected solutions (panels $a$ and $b$); the upper branch $TW1$- and $TW2$-connected solutions (panels $e$ and $f$); and the two successive $CS$-connected solutions (panels $c$ and $d$, and their symmetry-related counterparts) appear identical. The solutions mainly differ in the number of high- and low-speed streaks.
From lower to upper branches of both $TW1$- and $TW2$-connected two high-speed streaks grow at the solution fronts.
The $TW1$-connected solution is centered at a high-speed streak while in the center of the $TW2$-connected solution a low-speed streak is located.
Application of the rotational symmetry $\sigma_{xy}$ to $TW1$-, $TW2$- and $CS$-connected solutions creates symmetry-related invariant solutions. However, the branches of these solutions do not merge with the snakes-and-ladders bifurcation structures in the presence of wall-normal suction. 

\section{Conclusion}\label{conclusion}

We investigate the structural stability of the snakes-and-ladders bifurcation structures of invariant solutions of the three-dimensional Navier-Stokes equations in plane Couette flow. \cite{Salewski2019} show that adding a Coriolis force in PCF ($V_s=0$) that maintains the equivariance group, preserves the snakes-and-ladders bifurcation structure. Here, we show that applying a non-vanishing suction velocity that breaks the rotational symmetry of PCF modifies the snakes-and-ladders bifurcation structure. 
For non-zero but small suction velocity, the curve representing the branches of both travelling waves $TW1$ and $TW2=\sigma_{xy}TW1$ (at $V_s=0$) in a bifurcation diagram showing dissipation $D$ as a function of $Re$ split in two different snaking curves with alternating span of the oscillations in Reynolds number. The both equilibrium branches $EQ1$ and $EQ2=\sigma_z EQ1$ (at $V_s=0$) break up. The pitchfork bifurcations of the equilibrium branches that create rungs at $V_s=0$ are broken. The result are new solution branches formed from remainders of both the the broken equilibrium branches and the rungs. The returning state branches $RS$ connect one of the travelling wave branches to itself while the connecting state branches $CS$ connect $TW1$ and $TW2$.
Specific features of the bifurcation diagram including the symmetric splitting of the $TW1$ and $TW2$ branches and the ordering of $RS$ and $CS$ follow from symmetry arguments. 

At small but non-vanishing suction velocity, the modifications of the snakes-and-ladders bifurcation structure of three-dimensional solutions of Navier-Stokes equations in plane Couette flow are analogous to modifications of the snakes-and-ladders bifurcation structure observed within the one-dimensional Swift-Hohenberg equation with the qubic-quintic nonlinearity (SHE35), when a quadratic term is added. Introducing a quadratic term $\epsilon u^2$ with amplitude $\epsilon$ in the SHE35 breaks the odd symmetry of this system, $R_2 : x\rightarrow -x,\ u\rightarrow -u$ \citep{Houghton2011}. Breaking $R_2$ in the SHE35 modifies the snakes-and-ladders bifurcation structure in the following way. For non-vanishing $\epsilon$ both solution branches that are invariant under the action of the even symmetry, $R_1 : x\rightarrow -x,\ u\rightarrow u$ split into two distinct curves with alternating span of the oscillations in $r$ in an amplitude versus control parameter $r$ bifurcation diagram. This is analogous to the splitting of the $TW1$ and $TW2$ in the Navier-Stokes problem. As the $EQ$ branches in Navier-Stokes, the continuous snaking branches of solutions that are invariant under $R_2$ at $\epsilon=0$ break up into disconnected segments for non-zero $\epsilon$. Together with remainders of rung states at $\epsilon=0$ the remainders of these solutions form disconnected $s$- and $z$-shaped branches that connect to the split snaking solution branches that are invariant under $R_1$. The $s$- and the $z$-shaped solution branches in the modified snakes-and-ladders bifurcation structure of the one-dimensional SHE35 with the symmetry-breaking quadratic term thus behave analogous to the $CS$ and $RS$ solution branches in the modified snakes-and-ladders bifurcation structure of Navier-Stokes equations in PCF in the presence of non-vanishing but small wall-normal suction velocity. When the amplitude of the quadratic term $\epsilon$ is increased towards large values, the snakes-and-ladders bifurcation structure in the SHE35 eventually transforms into a modified bifurcation structure that resembles the snakes-and-ladders structure of the Swift-Hohenberg with the quadratic-cubic nonlinearity. This is not observed in the Navier-Stokes problem, where large suction causes the modified snakes-and-ladders bifurcation structure to disintegrate into disconnected branches of localised solutions. 
In conclusion, at small amplitudes, wall-normal suction in plane Couette flow and the symmetry-breaking quadratic term in the SHE35 yield analogous effects on the snakes-and-ladders bifurcation structure. At larger amplitudes, however, the effect of wall-normal suction in plane Couette flow and the symmetry-breaking term in SHE35 affect the bifurcation structure in significantly different ways.

Following the solution branches generated by suction back to regular plane Couette flow with $V_s=0$, additional disconnected branches of spatially localised invariant solutions of PCF are identified. Disconnected branches of solutions with varying width but matching front structure exists. The solutions thus share similarities with the snaking solutions but do not undergo homoclinic snaking themselves. While the snaking solutions only exist in a limited range of Reynolds number, the non-snaking solution branches span a much larger range of Reynolds numbers extending to those Reynolds numbers in which localised turbulent patterns are observed \citep{Barkley2005}. Consequently, the non-snaking localised invariant solutions identified here might be more relevant for supporting localised transitional turbulence than the previously known snaking branches. The dynamical relevance of the non-snaking localised solutions should be investigated in the future. 

\section*{Acknowledgements}
This work was supported by the Swiss National Science Foundation (SNSF) under grant no. 200021-160088. SA acknowledges support by the State Secretariat for Education, Research and Innovation SERI via the Swiss Government Excellence Scholarship.

\section*{Declaration of interests}
The authors report no conflict of interest.

\bibliographystyle{jfm_abbrv}
\bibliography{references}

\end{document}